\documentclass[12pt,epsfig]{iopart} 

\usepackage{iopams} 
\usepackage{epsf} 
\usepackage{epsfig}
\usepackage{color} 
\usepackage{psfrag}
\usepackage{verbatim} 
\usepackage{setstack} 
\usepackage{amsopn}
\usepackage{appendix} 
\usepackage{graphicx}
\DeclareMathOperator*{\sgn}{sign} \DeclareMathOperator*{\im}{Im}

\newcommand{\be}{\begin{equation}} \newcommand{\ee}{\end{equation}}
\newcommand{\reff}[1]{(\ref{#1})}
\renewcommand{\eref}[1]{Eq.~(\ref{#1})}

\begin{document}

\unitlength = 1mm \eqnobysec

\title[Critical dynamics with correlated noise]{Critical Langevin
dynamics of the $O(N)$-Ginzburg-Landau model with correlated noise}
\author{Julius Bonart$^1$, Leticia F. Cugliandolo$^1$ and Andrea
Gambassi$^{2,1}$}

\address{$^1$Laboratoire de Physique Th\'eorique et Hautes Energies,
Universit\'e Pierre et Marie Curie -- Paris VI, 4 Place Jussieu, Tour
13-14, 5\'eme \'etage, 75252 Paris Cedex 05, France\\ $^2$SISSA --
International School for Advanced Studies and INFN, via Bonomea 265,
34136 Trieste, Italy }

\begin{abstract} We use the perturbative renormalization group to
study classical stochastic processes with memory. We focus on the
generalized Langevin dynamics of the $\phi^4$ Ginzburg-Landau model
with additive noise, the correlations of which are local in space but
decay as a power-law with exponent $\alpha$ in time.
These correlations are assumed to be due to the coupling to an
equilibrium thermal bath.
We study both the equilibrium dynamics at the critical point and
quenches towards it,
deriving the corresponding scaling forms and the associated
\emph{equilibrium} and \emph{non-equilibrium} critical exponents
$\eta$, $\nu$, $z$ and $\theta$. We show that, while the first two
retain their equilibrium values independently of $\alpha$, the
non-Markovian character of the dynamics affects $z$ and $\theta$ for
$\alpha < \alpha_c(D, N)$ where $D$ is the spatial dimensionality, $N$
the number of components of the order parameter, and $\alpha_c(x,y)$ a
function which we determine at second order in $4-D$.
We analyze the dependence of the asymptotic fluctuation-dissipation
ratio on various parameters, including $\alpha$.  We discuss the
implications of our results for several physical situations.
\end{abstract}

\newpage

\section{Introduction}

For more than 30 years critical dynamics have been explored with field
theoretical
methods~\cite{Masiro73,deDo76,Bajawa76,Hahoma74,Hahoma76,Hamaho72,Hoha77,DDPe78,On02}.
A variety of dynamic models were introduced to describe the collective
evolution of systems close to critical points.  Among these, the most
common ones are the dynamics of non-conserved or conserved order
parameters which successfully describe the evolution of uniaxial
magnetic systems close to the Curie point or the dynamics of binary
alloys close to the demixing transition, respectively.  These problems
as well as many of their generalizations discussed
in~\cite{Hahoma74,Hahoma76,Hoha77,On02} are \emph{classical} in the
sense that their stochastic nature can be essentially ascribed to
thermal fluctuations. Therefore, as in the simpler diffusion
processes, the evolution of the interacting degrees of freedom ---
described via a field $\phi$ --- is modeled by a functional Langevin
equation in which the coupling to the environment is responsible for
both thermal fluctuations, encoded in a stochastic external noise, and
the friction.  If the environment, which acts as a thermal bath for
$\phi$, is in equilibrium at a certain temperature $\beta^{-1}$, the
space and time dependence of the friction coefficient and of the
correlations of the thermal fluctuations are related via the
fluctuation-dissipation theorem. The selected model bath determines
then the remaining functional form of the noise-noise correlation and
the usual choice is to take it to be delta-correlated in time which
corresponds to white noise.

In cases in which the initial conditions of the system are drawn from
an equilibrium Gibbs-Boltzmann distribution, or the system is allowed
to evolve for a sufficiently long time such that this distribution
function is reached, the space-time behavior of dynamic quantities is
characterized by scaling laws in which the usual static exponents
($\nu, \ \eta$ etc.) appear but a new critical exponent $z$ is needed
to relate the space and time dependencies.  Besides the analysis of
equilibrium dynamics, field-theoretical methods allow one to study the
\emph{non-equilibrium} dynamics after a sudden quench from a suitably
chosen initial condition to the critical point~\cite{Jascsc88}.  A
Gaussian distribution of the initial field configuration with zero
average and short-range correlations mimics a quench form the
disordered phase~\cite{Jascsc88}. A distribution with non-zero average
but still short-range correlations describes (in the case of a scalar
field $\phi$) a quench from the ordered state~\cite{cgk-06,cg-07}.  In
these non-equilibrium cases a new critical exponent --- usually
denoted by $\theta$ and called the ``initial slip exponent'' ---
characterizes the short-time behavior of the average order parameter
as well as of the correlation and response functions~\cite{Jascsc88}
(see, e.g.,~\cite{Jan92,cg-05,Ozit07} for summaries).

In all the studies mentioned above, the noise is assumed to have a
Gaussian distribution (as a consequence of the central limit theorem)
with no temporal correlations, i.e., to be Gaussian and
\emph{white}. However, the coupling to thermal reservoirs yields, in
general, non-Markovian Langevin equations in which the noise is
correlated in time and the friction coefficient has some
memory~\cite{Zw73,Ka73,We08}.  In many situations of practical
interest a full treatment of such a \emph{colored} noise is therefore
necessary. For example, the escape rate of particles confined within a
potential well crucially depends on the statistics of the thermal
bath~\cite{Hantabo90,Han94,Maweli87} as observed in the desorption of
molecules from a substrate undergoing a second-order phase transition
which effectively provides a colored noise for the stochastic dynamics
of the molecules (this phenomenon is called Hedvall effect and it was
studied theoretically in, e.g.,~\cite{SeDi-87}). Another important
instance is the stochastic Burgers modeling of turbulence where the
noise is correlated in both time and space~\cite{Mehwmazh89}. This
equation is, in addition, closely related to the Kardar-Parisi-Zhang
description of surface growth that was analyzed with spatially
correlated noise in, e.g.,~\cite{Jafrta99,Ka03} and references
therein.  The so-called fractional Brownian motion~\cite{Mandel68} is
actually defined by the non-Markovian nature of the process, which can
be traced back to the temporal correlations characterizing the thermal
noise. The physical circumstances in which a temporally correlated
noise arises are manifold including
polymer translocation through a nanopore~\cite{Zo09} or the effective
description of a tracer in a glassy medium~\cite{Po03}. A review of
the effects of colored noise in dynamical systems is given
in~\cite{Hanju95}. Last but not least, the environment fluctuations in
quantum dissipative systems give rise rather naturally to temporally
correlated contributions~\cite{We08}.

In the present study we explore the influence of colored noise on the
equilibrium and non-equilibrium critical dynamics of the $N$-component
$\phi^4$ Ginzburg-Landau model with $O(N)$ symmetry and non-conserved
order parameter. We focus on the case of noise correlations that decay
as a power-law in the time-lag, such that no time scale can be
associated to it at least for sufficiently long times.  To our
knowledge critical dynamics with such a colored noise have not been
theoretically investigated yet.  Our first aim is to present a general
equilibrium analysis that allows us to determine the classes of noise
which affect the critical relaxation. We use the perturbative
renormalization group (RG) technique to calculate the critical
exponents, which turn out to be modified for a certain type of noise
correlations only.  Our second aim is to understand the
non-equilibrium dynamics in the presence of such kind of colored
noise.  A short time after the critical quench, the dynamics of the
system can be described in terms of the non-equilibrium critical
exponent $\theta$ which depends on the properties of the colored
noise. In the critical coarsening regime we define an \emph{effective
temperature}~\cite{Cukupe97,Cuku00} $\beta^{-1}_\infty$ from the
long-time limit of the so-called fluctuation-dissipation ratio [c.f.,
\eref{eq:FDR} for its definition] which was proposed to be an
additional universal property of the non-equilibrium
dynamics~\cite{GL00}.  $\beta^{-1}_\infty$ was calculated for a
variety of critical processes with white
noise~\cite{cg-05,Coliza,Cu11}.
We show that $\beta^{-1}_\infty$ turns out to be affected by the color
of the noise and we compare our results to the corresponding ones for
the fractional Brownian motion~\cite{Po03}.

The paper is organized as follows. In Sec.~2 we present the model and
we obtain the associated dynamic action by using the path-integral
formalism~\cite{Masiro73,Bajawa76,deDo76,Hoha77,DDPe78,Arbicu10}. We
define the response and the correlation functions, we present their
scaling behavior and we define the critical exponents. In Sec.~3 we
focus on the equilibrium properties of the model, we calculate the
propagators in the frequency and momentum domain, presenting the
Feynman rules of the perturbation theory. This section contains
details on the perturbative RG analysis which allows us, on the one
hand, to determine the type of noise which alters the critical
dynamics and, on the other, to calculate the dynamical critical
exponent $z$ up to second order in $\epsilon=D_c-D$ where $D_c$ is the
upper critical dimensionality, which we find to be 4. In Sec.~4 we
focus on the non-equilibrium dynamics. We derive the critical scaling
behavior of the linear response and correlation functions and we
calculate the non-equilibrium critical exponent $\theta$ (the
``initial slip exponent'', see \cite{Jascsc88}) up to first order in
$4-D$. Using the scaling forms of the correlation and linear response
functions we deduce the effective temperature after a critical quench.
Finally, in Sec.~5 we summarize our results and we discuss several of
their possible applications together with plans for further
investigation.  The details of the calculations are reported in the
Appendices.

\section{The model}
\label{sec:model}

In $D$ spatial dimensions we consider an $N$-component non-conserved
order parameter $\vec{\phi}(\vec x)$ whose static behavior near the
critical point is governed by a Hamiltonian of the Ginzburg-Landau
type
\begin{equation}\label{equ1} \mathcal{H}[\vec{\phi}] = \int \rmd^Dx
\left[\frac{1}{2}(\vec\nabla\vec{\phi})^2+\frac{1}{2}r\vec{\phi}^2+\frac{g}{4!}\vec{\phi}^4\right]
.
\end{equation} $g$ is the strength of the non-linearity that drives
the phase
transition, $r$ is the control parameter for it, and the coefficient
in front of the elastic term $\propto (\vec\nabla\vec{\phi})^2 $ has
been absorbed in the definition of the field.  The time evolution
under purely dissipative dynamics (model A according to the standard
classification of~\cite{Hoha77}) in the presence of memory is
described by the Langevin equation
\begin{equation}\label{equ2} \int_{-T}^t \rmd
t'\;\Gamma(t-t')\partial_{t'}{\vec{\phi}}(\vec{x},t') +
\frac{\delta\mathcal{H}}{\delta\vec{\phi}(\vec{x},t)} =
\vec{\zeta}(\vec{x},t) ,
\end{equation} where $-T$ is the initial time of the process and
$\vec{\zeta}$ is a zero-mean Gaussian colored noise with
\begin{equation} \langle\zeta_i(\vec{x},t)\zeta_j(\vec{x}',t')\rangle
= \beta^{-1}\Gamma(t-t')\delta(\vec{x}-\vec{x}')\delta_{ij} .
  \label{eq:delta-corr}
\end{equation} $\Gamma$ is a positive and symmetric kernel, i.e.,
$\Gamma(t-t')=\Gamma(t'-t) > 0$.
Depending on the specific system of interest, in the minimal dynamical
model in Eq.~\reff{equ2} one might need to include an inertial term
$M\partial^2_t\phi$.  An order of magnitude analysis suggests that
such a term is relevant for times shorter than $M/\int\rmd t \
\Gamma(t)$, whereas the effects of inertia can be neglected for longer
times, indeed the regime we are interested in.  For this reason we
shall omit this term from the outset as long as the contribution of
friction does not vanish.
Note that the function $\Gamma$ determines both the noise-noise
correlation [Eq.~\reff{eq:delta-corr}] and the time-dependent retarded
friction coefficient [Eq.~\reff{equ2}] since we have assumed the
thermal bath (which is weakly coupled to the system) to be in
equilibrium at temperature $\beta^{-1}$ [we set $k_B = 1$]. The
Markovian examples of this dynamics are characterized by a
$\delta$-correlated ('white') noise, i.e., \be
\label{eq:white} \Gamma(t)=2\gamma_{\rm w}\delta(t), \ee where
$\gamma_{\rm w}$ is the friction coefficient. In this case \eref{equ2}
has been extensively studied both in and out of equilibrium, see,
e.g.,~\cite{Bajawa76,Jascsc88,cg-05,Caga02}. Such \emph{Ohmic}
dissipation is the simplest form of \emph{short-range correlated}
noise. It can be formally obtained as the limit $t_0\rightarrow 0$ of
the exponentially correlated Ornstein-Uhlenbeck (OU) process \be
\Gamma_{\rm OU}(t) = \frac{\gamma_{\rm OU}}{t_0}\rme^{-|t|/t_0}, \ee
where the finite characteristic relaxation time $t_0$ plays the role
of an internal scale. Under renormalization one expects the
exponentially correlated noise to become equivalent to a white
(delta-correlated) one, \eref{eq:white}, and the critical behavior of
the OU process be identical to the Markovian one. In the absence of an
internal scale, instead, there is no reason to expect a white noise
limit and the critical behavior might be affected. The simplest
example with no explicit time scale is
\begin{equation}\label{equ3} \Gamma(t) =
\frac{\gamma}{\Gamma_E(1-\alpha)} \ |t|^{-\alpha} \qquad\mbox{with}
\;\; \alpha >0.
\end{equation} We denote by $\Gamma_E$ the Euler $\Gamma$-function to
avoid confusion with the noise kernel $\Gamma$. For $\alpha>1$, i.e.,
\emph{super Ohmic dissipation}, expression~(\ref{equ3}) is not
integrable, unless a short-time cut-off and thus an internal scale is
introduced.  One can show that under naive scaling (introduced in
Sec.~3) the Fourier or Laplace transform of $\Gamma(t)$ generate a
white noise vertex that dominates over the colored noise part. Hence,
the appearance of a cut-off scale suggests the non relevance of the
colored noise for {\it super-Ohmic} dissipation, i.e., $\alpha>1$.
This statement will be made precise in the following. Instead, for
\emph{sub-Ohmic dissipation}, i.e., $\alpha<1$, the noise is truly
long-range correlated and its influence on the dynamics will turn out
to be non-trivial. The naive cross-over value between these two cases
is $\alpha=\alpha_c = 1$, that is white noise or \emph{Ohmic
dissipation}. In the presence of interactions we shall show that this
scenario is slightly modified, with the cross-over value
$\alpha_c(D,N)$ depending upon $D$ and $N$.

A functional-integral representation of the stochastic process,
Markovian or not, is better suited for an analytic treatment of
critical dynamics than the Langevin equation (\ref{equ2}).  In
particular, it allows one to express the average $\langle \cdots
\rangle_\zeta$ over the possible realizations of the noise $\vec
\zeta$ in Eq.~\reff{equ2} as a functional integral (which will be
denoted by $\langle \cdots\rangle$ in what follows) \be
\label{eq:MSR} \langle \cdots \rangle_\zeta = \int [\rmd \phi \rmd
\bar \phi] \; \cdots \; \rme^{-{\mathcal S}[\phi,\bar\phi]} \ee over
$\phi$ and an auxiliary field $\bar\phi$ with
$\mathcal{S} = \mathcal{S}_0 + \mathcal{S}_{int} - \ln
\mathcal{P}_{IC}$~\cite{Masiro73,deDo76,Bajawa76,Arbicu10}\footnote{In
the presence of colored noise no discretization problems arise, see,
e.g.,~\cite{Arbicu10}.},
\begin{eqnarray}\label{equ4} & \mathcal{S}_0 = \int \rmd^Dx
\int_{-T}^\infty \rmd t \ \overline{\phi}_i(\vec{x},t)
\left[\int_{-T}^t \rmd t'\;\Gamma(t-t')\partial_{t'}\phi_i(\vec{x},t')
+ (r-\nabla^2)\phi_i(\vec{x},t)\right] \nonumber\\ & \qquad
-\beta^{-1} \int \rmd^Dx\int_{-T}^\infty \rmd t \int_{-T}^t \rmd
t'\;\overline{\phi}_i(\vec{x},t)\Gamma(t-t')\overline{\phi}_i(\vec{x},t')
\end{eqnarray} and
\begin{equation}\label{equ5} \mathcal{S}_{int} =\int \rmd^Dx
\int_{-T}^\infty \rmd t
\;\frac{g}{3!}\overline{\phi}_i(\vec{x},t)\phi_i(\vec{x},t)\phi_j(\vec{x},t)\phi_j(\vec{x},t)
.
\end{equation} We used Einstein's convention of summation over
repeated indices. The zero-source functional integral is identical to
$1$ due to the normalization of the noise probability
distribution. $\mathcal{P}_{IC}[\vec{\phi}(\vec{x},-T)]$ is the
statistical weight of the initial condition. The auxiliary field
$\vec{\overline\phi}$\footnote{$\vec{\overline\phi}$ is purely
imaginary and it is sometimes written as $i\vec{\overline\phi}$ in the
literature.} is conjugated to an external perturbation $\vec h$, in
such a way that if $\mathcal{H}[\vec \phi,\vec h]=\mathcal{H}[\vec
\phi]-\vec \phi \cdot \vec h$, the linear response of the order
parameter to the field $\vec h$ is given by
\begin{equation}\label{equ6} R(\vec{x}-\vec{x}';t,t')\delta_{ij} =
\left.\frac{\delta\langle\phi_i(\vec{x},t)\rangle_{\vec h}}{\delta
h_j(\vec{x}',t')}\right|_{\vec h=\vec 0} =
\langle\phi_i(\vec{x},t)\overline{\phi}_j(\vec{x}',t')\rangle,
\label{eq:linear-resp}
\end{equation}
where $\langle \cdots \rangle_{\vec h}$ is the average over the
stochastic process in the presence of the external perturbation, i.e.,
Eq.~\reff{equ2} with $\mathcal{H} \mapsto \mathcal{H}[\vec \phi,\vec
h]$.
The response function is causal irrespectively of the noise statistics
and the Jacobian of the transformation of variables from $\vec \zeta$
to $\vec \phi$ which allows us to write the average over the
stochastic process as in Eq.~\reff{eq:MSR} is also a factor with no
consequences~\cite{Arbicu10}.
In addition to the (linear) response function, we shall consider below
the correlation function of the order parameter, defined by \be
\label{eq:corr-gen} C(\vec{x}-\vec{x}',t,t')\delta_{ij} =
\langle\phi_i(\vec{x},t)\phi_j(\vec{x}',t')\rangle \ee where we
assumed translational invariance in space.
The action ${\mathcal S}_0 + {\mathcal S}_{int}$ is the sum of two
contributions each one made of several terms.  The part with density
$\bar\phi_i \delta{\mathcal H}/\delta\phi_i$ represents the
deterministic dynamics whereas the remaining part is due to 
the coupling to the bath. The latter consists of the
friction term and the noise-noise correlation and both involve the
kernel $\Gamma$.  In this formulation the problem is recast in the
form of a field theory in $D+1$ dimensions with two vector fields, the
analysis of which can be done via standard field-theoretical tools,
such as the renormalization group (RG) approach that we shall use
below.

Since, in general, there is no tractable Fokker-Planck equation for
the non-Markov stochastic processes we are presently interested in,
the usual and relatively simple proof of equilibration explained in,
e.g.,~\cite{Zi96,Pa} for the white-noise problem does not apply.
However, we recall here that \eref{equ2} is an effective description
of the dynamics of a classical system with Hamiltonian ${\mathcal H}'$
which is weakly and linearly coupled to a (large) equilibrium bath of
harmonic oscillators at temperature $\beta^{-1}$, acting as a source
of the stochastic noise $\vec\zeta$ effectively induced by such a
coupling. Indeed, the temperature that characterizes the correlations
of the noise in \eref{eq:delta-corr} is $\beta^{-1}$, whereas the
distribution of the frequencies of the harmonic oscillators within the
bath determines the functional form of $\Gamma$. In addition, $\Gamma$
appears in \eref{equ2} and \eref{eq:delta-corr} in such a way to
ensure the fluctuation-dissipation theorem for the bath variables
[see, c.f., \eref{equ8}].
As a result, even with this effective non-Markov dynamics the system
should still lose memory of its initial condition and equilibrate with
the equilibrium bath of oscillators, resulting in a canonical
distribution $\rme^{-\beta {\cal H}[\vec \phi]}/{\cal Z}(\beta)$ of
one-time quantities at sufficiently long times (possibly divergent
with the system size) where ${\cal Z}(\beta)$ is the partition
function and ${\cal H}$ differs from ${\cal H}'$ by a term which is
quadratic in the relevant degrees of freedom (see,
e.g.,~\cite{We08,Cu-02} for details).
The asymptotic critical \emph{equilibrium dynamics} is expected to be
described by the limit $T\rightarrow\infty$ of the action in which one
neglects the specific distribution ${\mathcal P}_{IC}$ of the initial
conditions that in any case should be forgotten dynamically.  Since we
shall be interested in the critical dynamics, we set $\beta=\beta_c$
and we absorb this constant into a redefinition of the fields and of
the coupling constant $g$.
In equilibrium the response and the correlation functions are
invariant under time translations, i.e., $R(\vec{x},t,t') =
R(\vec{x},t-t')$ [see \eref{eq:linear-resp}] and $C(\vec{x},t,t') =
C(\vec{x},t-t')$ [\eref{eq:corr-gen}] , and they are related 
to each other by the fluctuation-dissipation
theorem (FDT) that reads $R(\vec{x},t) =
-\beta\partial_tC(\vec{x},t)\Theta(t)$, where $t$ represents the time
delay, $\Theta(t\le 0)=0$ and $\Theta(t> 0)=1$ \footnote{Note that the
It\^o prescription of the Langevin equation (\ref{equ2}) implies
$\Theta(0) = 0$ in the stochastic path integral description
\cite{Zi96,Bajawa76}.}, and which is completely independent of the
specific characteristics of the system and the bath apart from its
temperature. (A proof of this relation for generic colored noise
Langevin dynamics can be found in~\cite{Arbicu10}.) Once the latter
has been absorbed in the redefinition of $\phi_i$ and $g$ the FDT
becomes
\begin{equation}\label{equ8} R(\vec{x},t) =
-\partial_tC(\vec{x},t)\Theta(t) ,
\end{equation} and this is the form that we shall use in our
calculations.  Moreover, the time-dependent correlation is invariant
under time-reversal, i.e., $C(\vec{x},t)=C(\vec{x},-t)$.

\emph{Non-equilibrium dynamics}, instead, can be studied by leaving
$T$ finite and by making the initial distribution $\mathcal{P}_{IC}$
explicit~\cite{Jascsc88,cg-05}. A typical choice is a Gaussian weight
in which case $\beta_c$ can still be absorbed into a redefinition of
the fields and $g$. Stationarity is lost out of equilibrium and
correlation and linear response functions depend on all times involved
in their definitions ($t$ and $t'$ in Eqs.~\reff{eq:linear-resp} and
\reff{eq:corr-gen}). Moreover, the FDT is no longer
valid~\cite{cg-05,Coliza,Cu11}.

 In addition to $R$ defined in \eref{eq:linear-resp} and $C$ defined
in \eref{eq:corr-gen}, one can construct the quadratic correlator $
\langle \overline\phi_i(\vec x,t) \overline\phi_j(\vec x',t') \rangle$
which, independently of the color of the noise, vanishes identically
due to causality.

\subsection{Scaling}

In the case of stochastic dynamics with white noise, a systematic RG
analysis confirms the phenomenological scaling behavior of the linear
response and correlation functions both for $T\to\infty$ and $T$
finite corresponding, respectively, to equilibrium and non-equilibrium
relaxational dynamics.
In terms of the equilibrium correlation length $\xi_{\rm eq}\simeq
|r-r_c|^{-\nu}$, where $r_c$ is the critical value of the parameter
$r$ in Eq.~\reff{equ1}, and of a dynamic growing length $\xi(t)\simeq
t^{1/z}$, one expects~\cite{Jascsc88,cg-05}
\be
\label{scalingR} R(\vec{p},t,t') = p^{-2+\eta+z} \
[\xi(t)/\xi(t')]^{z\theta} \ f_R(p\xi_{\rm eq},\xi(t)/\xi_{\rm
eq},\xi(t')/\xi(t)) \ee and \be
\label{scalingC} C(\vec{p},t,t') = p^{-2+\eta} \
[\xi(t)/\xi(t')]^{z(\theta-\hat{\alpha})} \ f_C(p\xi_{\rm
eq},\xi(t)/\xi_{\rm eq},\xi(t')/\xi(t)) , \ee for the Fourier
transform in space of $R(\vec{x},t,t')$ and $C(\vec{x},t,t')$,
respectively, with the white noise value $\hat{\alpha}=1$
\cite{Jascsc88}. In the previous expressions, $\nu$ is the standard
static critical exponent associated with the correlation length, $z$
is the dynamic critical exponent which characterizes the different
scaling behavior of space and time, whereas $\eta$ is the static
anomalous dimension of the field $\phi$ and it controls the power-law
spatial decay of the static correlation function. $\theta$ is the
so-called initial-slip exponent \cite{Jascsc88,Jan92,cg-05} that
accounts for the effects of the initial condition in the case of
finite $T$. It is a novel universal quantity if the relaxation occurs
from a disordered initial state, whereas it is related to known
equilibrium exponents if the initial state has a non-vanishing average
value of the order parameter \cite{cgk-06,cg-07}. In \eref{scalingR}
and \eref{scalingC} $f_{R,C}$ are scaling functions which become
universal after the introduction of proper normalization. Equilibrium
dynamical scaling is recovered in the limiting case $\xi(t')\simeq
\xi(t) \gg \xi_{\rm eq}$ (i.e., in the limit of long times $t$, $t'$
with finite $t-t'$), whereas aging phenomena are expected to emerge
for $\xi(t),\xi(t') \ll \xi_{\rm eq}$ and, in particular, right at the
critical point $r=r_c$.  In the presence of specific instances of
correlated noise we expect the scaling behavior in \eref{scalingR} and
\eref{scalingC} to be modified both as far as the exponents and the
scaling functions are concerned.  The changes appear at the level of
the Gaussian theory and non-trivial effects survive in the presence of
interactions for certain noise correlations, as we shall explain in
Secs.~3 and 4.


\section{Equilibrium dynamics}
\label{sec:eq}

According to the interpretation of the Langevin dynamics in
\eref{equ2} as resulting from the coupling to an equilibrium thermal
bath, after a sufficiently long time the system is expected to relax
towards an equilibrium state characterized by the effective
Hamiltonian ${\mathcal H}$, i.e., by the static $\phi^4$-theory.
This relaxation occurs generically and for arbitrary initial
conditions as long as the asymptotic values of the control parameters
of the system ($r$ in the case we are concerned with) imply for $
{\mathcal H}$ neither a spontaneous symmetry breaking nor criticality,
which would indeed provide instances of \emph{aging} (see, e.g.,
\cite{Cu-02}).
However, the existence of a wide region of parameter space ($r>0$) for
which equilibration occurs, allows us to conclude that all static
properties of a theory with effective Hamiltonian ${\mathcal H}$ carry
over to the dynamic field-theoretical action ${\mathcal S}$
[see~\eref{equ4} and \eref{equ5}] which generates the dynamic
correlation functions and therefore the static ones as a special
case. The upper critical dimensionality $D_c$ above which the Gaussian
theory becomes exact is therefore the same as in the $\phi^4$ theory,
i.e., $D_c=4$ (see, e.g., \cite{Zi96}). Analogously, the same applies
to the static exponents $\nu$ and $\eta$.  In this section we show how
this arises within perturbation theory. In particular, we determine
the conditions under which the critical dynamics is modified by the
colored part of the noise with special focus on the emergence of a
cross-over line $\alpha_c(D,N)$ which bounds the region within which
the dynamic exponent $z$ is affected by the color of the noise. We
calculate this exponent in the white and colored noise cases.

\subsection{Gaussian theory}

\label{sec:Gaux}

In the $T\to\infty$ limit the Gaussian part of the action
$\mathcal{S}_0$ can be diagonalized via a Fourier transform of the
fields defined in Eq.~(\ref{equA101}). One obtains \be \mathcal{S}_0 =
\frac{1}{2}
\int\frac{\rmd\omega}{2\pi}\frac{\rmd\omega'}{2\pi}\int\frac{\rmd^Dp}{(2\pi)^D}\frac{\rmd^Dp'}{(2\pi)^D}\;
\vec{\varphi}^T(\vec{p},\omega)\mathcal{C}(\vec{p},\omega;\vec{p'},\omega')\vec{\varphi}(\vec{p'},\omega'),
\label{S0Gaux} \ee where we used a vector notation $\vec{\varphi} =
(\vec
\phi(\vec{p},\omega)\,,\,\vec{\overline{\phi}}(\vec{p},\omega))^T$ for
the $2N$-component field $\vec{\varphi}$ and we introduced the
correlation matrix \be \mathcal{C}
=\delta_{ij}\delta(\vec{p}+\vec{p}')\delta(\omega+\omega')\left( \begin{array}{ccc}
0 & i\omega\Gamma_{i\omega}+(p^2+r)\\
-i\omega\Gamma^*_{i\omega}+(p^2+r) &
-(\Gamma_{i\omega}+\Gamma^*_{i\omega})\ \end{array} \right) .
\label{eq:corrMatrix} \ee Here and in what follows we denote a
function and its Fourier transform with the same symbol, the
difference being made clear by their arguments. In
\eref{eq:corrMatrix} $\Gamma_{i\omega}$ stands for the Fourier
transform of $\Theta (t) \Gamma(t)$ [the $\Theta(t)$ factor is a
consequence of the causal structure of \eref{equ2}]. As usual, $^*$
denotes the complex conjugate. For the colored noise in \eref{equ3}
one finds
\begin{equation}\label{equN} \Gamma_{i\omega}=
\gamma|\omega|^{\alpha-1}\left[\sin(\pi\alpha/2) -
i\sgn(\omega)\cos(\pi\alpha/2)\right] + \gamma_{\rm w}.
\end{equation}
[Note that for $\alpha>1$ a short-time cut-off has to be introduced in
order to transform \eref{equ3}.  However, the dynamic properties we
are presently interested in are determined by the leading behavior at
small $\omega$, which is not affected by the introduction of such a
cut-off and is correctly captured by \eref{equN}. Accordingly, we
shall use this form irrespectively of the value of $\alpha$.]
In this expression we have added a supplementary \emph{white-noise
vertex} $\gamma_{\rm w}$ for reasons that will become clear in the
following [note that the cut-off that has to be introduced in order to
make \eref{equ3} integrable for $ \alpha>1$ effectively leads to this
supplementary white-noise vertex].  The propagators\footnote{In what
follows we denote the response and correlation function by $R$ and
$C$, respectively.  The various propagators and quantities within the
Gaussian approximation are denoted by the subscript $_0$.} are deduced
by inverting $\mathcal{C}$:
\begin{equation}\label{equR} R_0(\vec{p},\omega) \delta_{ij} = \langle
\phi_i(\vec{p},\omega)\overline{\phi}_j(-\vec{p},-\omega) \rangle =
\frac{1}{i\omega\Gamma_{i\omega} + p^2 + r} \ \delta_{ij}
\end{equation} and
\begin{eqnarray}
\label{equC} C_0(\vec{p},\omega) \delta_{ij} &=\langle
\phi_i(\vec{p},\omega)\phi_j(-\vec{p},-\omega)\rangle\nonumber\\ &=
\frac{\Gamma_{i\omega}+\Gamma^*_{i\omega}}{\omega^2\Gamma_{i\omega}\Gamma^*_{i\omega}+i\omega(p^2+r)(\Gamma_{i\omega}-\Gamma^*_{i\omega})+(p^2+r)^2}
\ \delta_{ij}.
\end{eqnarray} By construction they satisfy the FDT [see \eref{equ8}]
that in the frequency domain reads: \be 2i\im R_0(\vec{p},\omega) =
-i\omega C_0(\vec{p},\omega).
\label{eq:FDTomega} \ee We recall that we absorbed the temperature
$\beta^{-1}$ in a redefinition of the fields and the coupling constant
$g$, and that $C_0(\vec p,\omega)$ is a real function.

The static correlation function $C_0(\vec{p},t=0)$ can be obtained by
integrating \eref{equC} over the frequency $\omega$ and, as expected,
the result agrees with the static Gaussian correlation that one would
infer from the Hamiltonian ${\mathcal H}$ [see, c.f., the calculation
leading to \eref{equA3b}]. Consequently, the static critical exponents
$\nu$ and $\eta$ are not modified at this order by the dynamics and
they take the Gaussian values $\nu_0=1/2$ and $\eta_0=0$,
respectively.

We anticipate here that in Sec.~\ref{sec:neqpr}, while discussing the
non-equilibrium dynamics of the present model, we consider the Laplace
transform [see \eref{equA102}] of \eref{equ2} with $g=0$ and the
colored noise given in \eref{equ3} (i.e., with $\gamma_{\rm
w}=0$). This allows us to determine the Laplace transform of the
response function $R_0$, formally obtained by replacing $i\omega$ with
$\lambda$ in \eref{equR}; compare Eqs.~\reff{equA101}
and~\reff{equA102}. This transform can be inverted to a form given in
terms of the so-called generalized Mittag-Leffler functions
$E_{\alpha,\beta}$ defined in \eref{def:genML} and provides a closed
expression for $R_0(\vec{p},t)$: \be R_0(\vec{p},t) = \Theta(t)
\frac{t^{\alpha-1}}{\gamma} E_{\alpha,\alpha}(-At^\alpha/\gamma) ,
\label{eq:R0-ML-ant} \ee where $A\equiv p^{2}+r$. The equilibrium
correlation function $C_0$ is readily determined from this expression
via the fluctuation-dissipation theorem \eref{equ8} (see, c.f.,
App.~\ref{Sec:Mittag} for details): \be C_0(\vec{p},t) = \frac{1}{A}
E_\alpha(-A|t|^\alpha/\gamma)
\label{eq:C0-ML-ant} \ee where $E_\alpha(z)\equiv E_{\alpha,1}(z)$.
\begin{figure}[h]
  \begin{center}
 \begin{tabular}{cc}
  \includegraphics[width=0.47\textwidth]{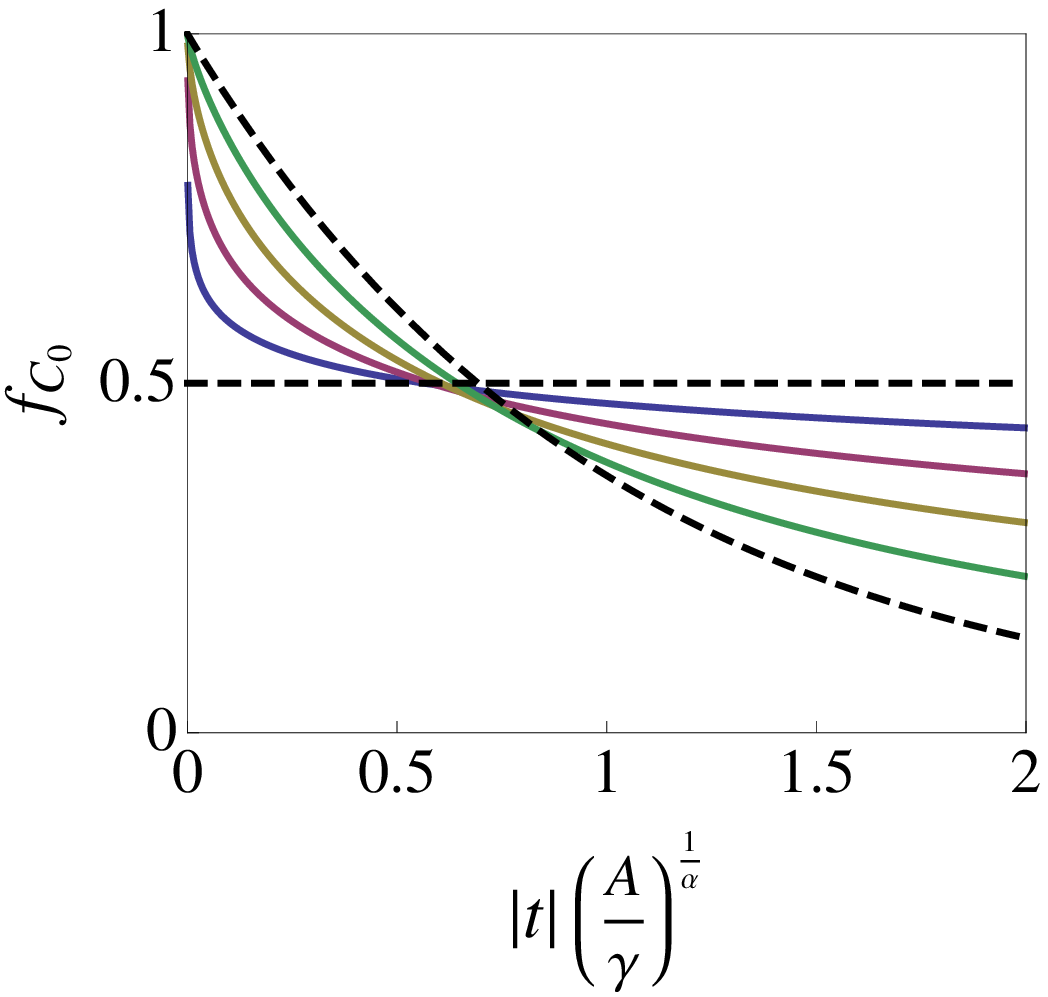} &
  \includegraphics[width=0.482\textwidth]{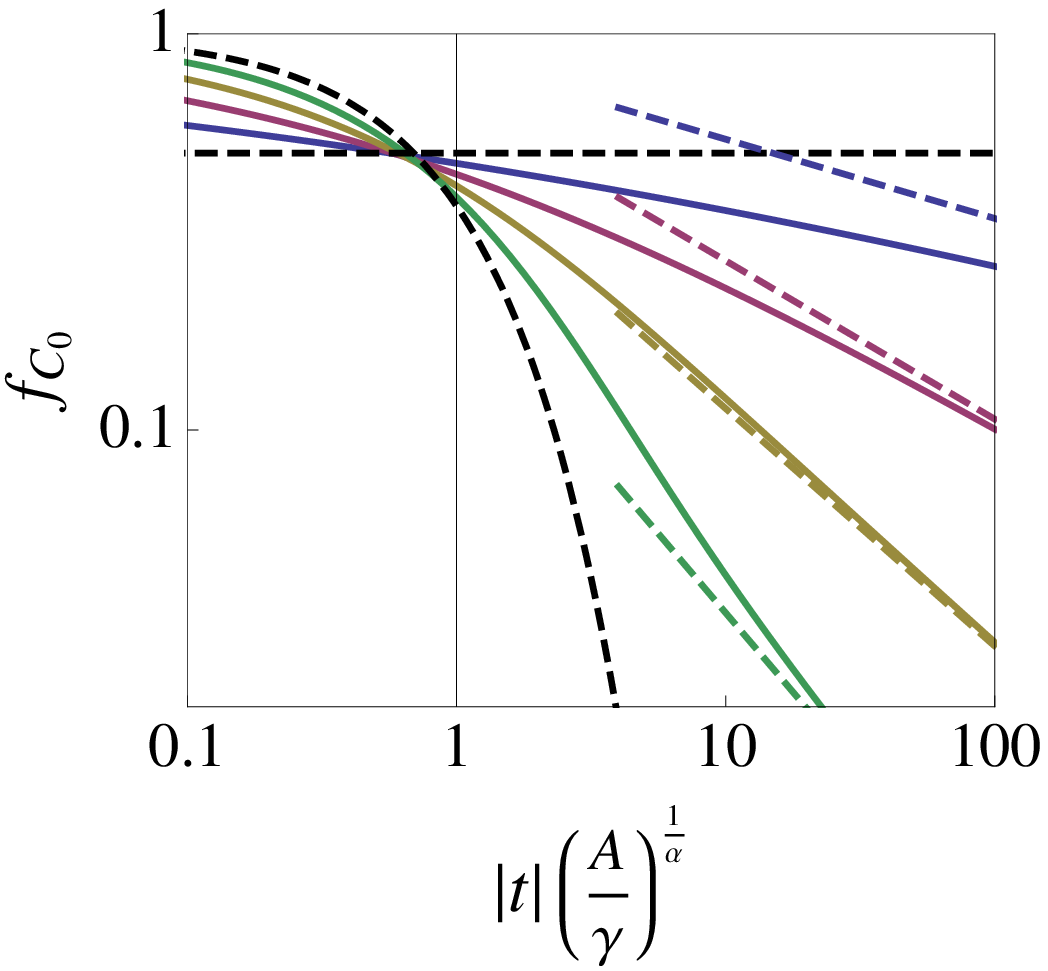} \\[2mm]
(a) & (b)
  \end{tabular}
  \end{center}
\caption{Scaling function $A C_0$ of the Gaussian correlation $C_0$ in
equilibrium ($T\to\infty$) as a function of $u\equiv
|t|(A/\gamma)^{1/\alpha}$ for various values of $\alpha$, with $A =
\vec{p}^{\,2}+r$.  (a) The horizontal dashed line corresponds to the
limit $\alpha \rightarrow 0$, whereas the other dashed line points out
the purely exponential behavior $\rme^{-u}$, which is recovered for
$\alpha = 1$. The solid lines, instead, correspond to $\alpha=0.2$,
0.4, 0.6, and 0.8, from bottom to top at small $u$ and in the reverse
order at large $u$. (b) Log-log plot of the curves shown in panel (a)
compared to their corresponding leading asymptotic algebraic behavior
inferred from \reff{eq:ML-asy}, which are indicated as (straight)
dashed lines.}
 \label{fig-corrfunc}
\end{figure}
In Fig.~\ref{fig-corrfunc}(a) we plot $A C_0$ as a function of the
(dimensionless) scaling variable $u\equiv |t|(A/\gamma)^{1/\alpha}$
associated with time $t$. For $\alpha\rightarrow 1$ one recovers the
purely exponential dependence $\rme^{-u}$ (indicated by the decreasing
dashed curve in Fig.~\ref{fig-corrfunc}) which characterizes the case
of white noise. As $\alpha$ decreases, instead, the correlation
function displays a faster initial drop followed by a slower decay at
large values of $u$. Indeed, taking into account the known asymptotic
behavior of the Mittag-Leffler functions [c.f., \eref{eq:ML-asy}],
these curves decay algebraically as $\sim 1/
[\Gamma_E(1-\alpha)u^{\alpha}]$ for $u\rightarrow\infty$. In panel (b)
of Fig.~\ref{fig-corrfunc} we use a log-log-scale to compare the
curves shown in panel (a) with their corresponding leading asymptotic
algebraic decays, indicated by the straight dashed curves for
$u\gtrsim 5$. As $\alpha \rightarrow 0$ the approximation provided by
the leading term of the asymptotic expansion becomes less accurate in
this time span and one needs to go to longer times to reach the
asymptotic regime. The curves in Fig.~\ref{fig-corrfunc} clearly
illustrate the crossover between an exponential and an algebraic
asymptotic behavior of the correlation function as $\alpha$ decreases
below the value $\alpha=1$.

For the generic case of the noise in \eref{equN} with $\gamma,
\gamma_{\rm w}\neq 0$, the propagators $R_0$ and $C_0$ do not have a
simple analytic form in the time domain, in contrast to the familiar
exponential relaxation which characterizes the case with white noise
($\gamma =0$, $\gamma_{\rm w}\neq0$) briefly recalled in
Eqs.~\reff{equA1} and \reff{equA2} and to the purely colored case
discussed in the previous paragraph ($\gamma \neq 0$ and $\gamma_{\rm
w}=0$).
In spite of this difficulty, the Gaussian value $z_0$ of the dynamic
exponent $z$ can be determined by comparing the scaling of the first
two terms in the denominator of $R_0$ for small $\omega$ and $p$ since
one expects $\omega \sim p^z$ from the definition of $z$ (see,
e.g.,~\cite{Zi96}).  First we note that for small $\omega$,
\eref{equN} scales as $\Gamma_{i\omega} \sim |\omega|^{\alpha-1}$ for
$\alpha < 1$, whereas $\Gamma_{i\omega} \sim 1$ for $\alpha > 1$: in
the former case the effect of the colored part of the vertex is
dominant, whereas in the latter the contribution of the white noise
($\propto\gamma_{\rm w}$) dominates.  As a result, from the scaling
$\omega \Gamma_{i\omega} \sim p^2$ one can read the Gaussian value
$z_0$ of the dynamic exponent:
\begin{equation}
\label{equZa} z_0 = \cases{ z_0^{\rm (col)} = 2/\alpha &for
$\alpha<1$,\\ z_0^{\rm (w)} = 2 &for $\alpha\geq 1$.}
\end{equation} A similar effect is observed in diffusion processes
with colored noise, the so-called fractional Brownian
motion~\cite{Mandel68}.  The particle's displacement is stationary and
characterized by an $\alpha$--dependent exponent which is called
\emph{Hurst exponent} in this context.

By rescaling the momentum $p$ and frequency $\omega$ according to
$p\mapsto b^{-1}p$ and $\omega\mapsto b^{-z}\omega$ with $b$ the
\emph{scaling parameter} of the RG flow, one deduces the Gaussian
scaling behavior of the response and the correlation propagator. We
infer from Eqs.~\reff{equR} and \reff{equC} that \be
b^{-2}R_0(b^{-1}\vec{p},b^{-z_0}\omega; r, \gamma,\gamma_{\rm w}) =
R_0(\vec{p},\omega; b^2r, b^{2-\alpha z_0}\gamma,b^{2- z_0}\gamma_{\rm
w}),
\label{eq:scalRGaux} \ee with a similar expression for $C_0$, where
the prefactor $b^{-2}$ on the left-hand side (lhs) is replaced by
$b^{-2-z}$.  As anticipated, one can identify two asymptotically
scale-invariant behaviors (the so-called Gaussian fixed-points in the
parameter space) as the Gaussian critical point $r=0$ is approached.
They correspond to $P\equiv(\gamma_{\rm w}=0,\gamma\neq 0)$ for
$\alpha<1$ and $P_{\rm w}\equiv (\gamma_{\rm w}\neq 0,\gamma=0)$ for
$\alpha\geq 1$, i.e., to the cases in which either the colored or the
white noise is relevant. The latter reduces to the standard Model A
dynamics~\cite{Zi96}. In order for $P$ and $P_{\rm w}$ to be fixed
points, it is necessary that the corresponding non-vanishing coupling
strengths, either $\gamma$ or $\gamma_{\rm w}$, are constant under
renormalization which, as expected from \eref{equZa}, implies
$z=z_0^{\rm (col)} = 2/\alpha$ for $\alpha<1$ ($P$) and $z=z_0^{\rm
(w)} = 2$ for $\alpha\geq 1$ $(P_{\rm w})$.

In order for the action ${\mathcal S}_0$ to be invariant under the
momentum and frequency rescaling discussed above, one has to rescale
the fields $\phi_i$ and $\overline\phi_i$ as $\phi_i(b^{-1}\vec
p,b^{-z_0}\omega) \mapsto b^{d_\phi} \phi_i(\vec p,\omega)$ and
$\overline\phi_i(b^{-1}\vec p,b^{-z_0}\omega) \mapsto
b^{d_{\overline\phi}} \overline\phi_i(\vec p,\omega)$ where $d_{\phi}$
and $d_{\overline\phi}$ are the so-called scaling dimensions of the
fields $\vec \phi$ and $\vec {\overline\phi}$, respectively, in the
$(\vec{p},\omega)$-domain. (Below we shall introduce the scaling
dimensions of the fields in the time-domain; in order to keep the
notation as simple as possible we do not include an additional
subscript to distinguish the two cases but we explain in the text
which one we use in each case.) The latter take the Gaussian values
\begin{eqnarray} d_{\phi,0} &= (D + 2)/2 + z_0, \label{equZb} \\
d_{\overline\phi,0} &= (D +2)/2. \label{equZc}
\end{eqnarray}
In the white-noise case $z_0=2$ we recover the standard scaling
dimensions of Model A critical dynamics \cite{Zi96}.  As far as the
transformation properties of the propagators under these rescalings
are concerned we have $b^{-2d_\phi+D+z_0}C_0(b^{-1}\vec
p,b^{-z_0}\omega;\ldots) = C_0(\vec p,\omega;\ldots)$ and
$b^{-d_\phi-d_{\overline\phi}+D+z_0}R_0(b^{-1}\vec
p,b^{-z_0}\omega;\ldots) = R_0(\vec p,\omega;\ldots)$ where the factor
$b^{D+z_0}$ comes from the $\delta$-function which guarantees the
conservation of momenta and frequencies. We have not specified the
scaling of the parameters $r$, $\gamma$ and $\gamma_{\rm w}$ to
lighten the notation.  By comparing with the scaling behavior of the
Gaussian response in \eref{eq:scalRGaux} and of the correlation
function, one confirms the Gaussian values \reff{equZb} and
\reff{equZc} for the dimensions $d_\phi$ and $d_{\bar\phi}$,
respectively.

In \eref{equN} we added to the colored-noise vertex associated with
\eref{equ3} a white-noise contribution proportional to $\gamma_{\rm
w}$ for the purpose of highlighting the emergence of the two distinct
Gaussian fixed points $P$ and $P_{\rm w}$. As we shall show below such
a white-noise contribution is anyhow generated under the RG flow as
soon as one accounts for the effect of non-Gaussian fluctuations
(i.e., $g\neq 0$) on the Gaussian fixed-point $P = (\gamma\neq
0,\gamma_{\rm w}=0)$ with colored noise alone.

\subsection{The interaction part}

The interaction part of the action reads
\begin{eqnarray*} \mathcal{S}_{int}&=&
\int\frac{\rmd\omega}{2\pi}\frac{\rmd\omega'}{2\pi}\frac{\rmd\omega''}{2\pi}\frac{\rmd^Dp}{(2\pi)^D}
\frac{\rmd^Dp'}{(2\pi)^D}\frac{\rmd^Dp''}{(2\pi)^D}\;
\frac{g}{3!}\overline{\phi}(-\vec{p}-\vec{p'}-\vec{p''},-\omega-\omega'-\omega'')
\nonumber\\ && \qquad \qquad \qquad \qquad \times
\phi(\vec{p},\omega)\phi(\vec{p'},\omega')\phi(\vec{p''},\omega'')
\end{eqnarray*} in the frequency and momentum domain. Under the naive
scaling with Eqs.~\reff{equZa}, \reff{equZb}, and \reff{equZc} one
easily obtains the scaling of the coupling constant: $g\rightarrow
b^{4-D}g$. The upper critical dimension is thus $D_c = 4$
independently of $\alpha$ and the effects of fluctuations beyond
mean-field can be accounted for by using a standard perturbative
expansion in terms of $\epsilon = 4-D$.

In the presence of the interaction ${\mathcal S}_{int}$, the scaling
dimension of the fields and the coupling constants are altered. In
addition, we shall show that the crossover value $\alpha_c=1$, which
separates the colored-noise-dominated case from the
white-noise-dominated one, acquires a dependance on $D$, thus dividing
the $(\alpha,D)$-plane (for $N$ fixed) in two distinct regions, each
one with different scaling properties. Under a RG flow with scaling
parameter $b>1$ the noise strengths $\gamma$ and $\gamma_{\rm w}$
scale as
\begin{eqnarray} \gamma &\mapsto& b^{2-\alpha z_0+\alpha\eta_\gamma}\
\gamma \label{scalinggamma}, \\ \gamma_{\rm w} &\mapsto& b^{2-z_0 +
\eta_{\rm w}}\ \gamma_{\rm w} \label{scalingwhite},
\end{eqnarray} which generalize the corresponding Gaussian scaling
behavior of these parameters --- encoded in \eref{eq:scalRGaux} ---
via the introduction of suitable anomalous dimensions $\eta_\gamma$
and $\eta_{\rm w}$ of $\gamma$ and $\gamma_{\rm w}$, respectively.
These anomalous dimensions $\eta_\gamma$ and $\eta_{\rm w}$  
determine the corrections to the Gaussian value $z_0$ of the dynamical
exponent $z$ and the crossover value $\alpha_c$ which separates the
different regions in the $(\alpha,D)$-plane. Indeed, let $l$ be a
length scale and $\tau$ be a time scale. Dimensional analysis implies
$t\sim\tau$ and $x\sim l$. From \eref{equR} we infer that $\gamma\sim
\tau^\alpha/l^2$ and $\gamma_{\rm w}\sim \tau/l^2$. Consider the case
in which the colored noise dominates, which corresponds to having 
$2-\alpha z_0+\alpha\eta_\gamma>2-z_0 + \eta_{\rm
w}$ in terms of the dimensions of the noise strengths [see
Eqs.~\reff{scalinggamma} and \reff{scalingwhite}] with $z_0 =
2/\alpha$. By choosing $\tau^\alpha = l^2\gamma$ we have $t \sim
l^{2/\alpha}\gamma^{1/\alpha}$. Therefore, under an RG flow with
$l\mapsto bl$ $(b>1)$ we deduce from \eref{scalinggamma} that $t \sim
b^{2/\alpha + \eta_\gamma}\ l^{2/\alpha}\gamma^{1/\alpha}$. On the
other hand, by noting that the dynamic exponent $z$ is defined through
$t \to b^{z} t$ we can readily identify the dynamic exponent $z =
2/\alpha + \eta_\gamma$ in terms of $\eta_\gamma$. In the
white-noise-dominated case we choose $\tau = l^2\gamma_{\rm w}$ and a
similar argument yields the white-noise result $z = 2 + \eta_{\rm
w}$. In short,
  \begin{equation} z = \cases{ 2+\eta_{\rm w} \qquad\qquad \mbox{for}
\;\; \alpha > \alpha_c(D,N) , \\ 2/\alpha+\eta_{\gamma} \qquad\;\;\;\,
\mbox{for} \;\; \alpha < \alpha_c(D,N), }
 \end{equation} and therefore one needs to calculate $\eta_{\rm w}$
and $\eta_\gamma$ in order to determine $z$.

In the presence of non-Gaussian fluctuations, the scaling dimensions
$d_{\phi} = d_{0,\phi}-z_0-\eta/2$ and $d_{\overline\phi} =
d_{0,\overline\phi}-z_0-\overline\eta/2$ in the $(\vec p,t)$-domain of
the fields $\phi$ and $\bar\phi$, respectively, differ from their
Gaussian values by the corresponding anomalous dimensions $\eta$ and
$\overline\eta$ (the extra $-z_0$ comes from the conversion of
$d_{0,\phi}$ and $d_{0,\overline\phi}$ from the frequency to the time
domain).  In order to determine the resulting scaling in the $(\vec
p,\omega)$-domain one has to take into account the integral over time
that carries a dimension $z$ (which differs from the Gaussian value
$z_0$); therefore
\begin{eqnarray} &\phi_i(b^{-1}\vec{p},b^{-z}\omega)\mapsto
b^{d_\phi+z-z_0}\ \phi_i(\vec{p},\omega) = b^{D/2+1+z-\eta/2}\
\phi_i(\vec{p},\omega),
\label{equ110}\\ &\overline{\phi}_i(b^{-1}\vec{p},b^{-z}\omega)\mapsto
b^{d_{\overline\phi}+z-z_0} \ \overline{\phi}_i(\vec{p},\omega) =
b^{D/2+1+z-z_0-\overline\eta/2}\ \phi_i(\vec{p},\omega)\label{equ111}
.
\end{eqnarray}

The FDT implies a relation between $\eta_{\gamma}$, $\eta_{\rm w}$,
$\eta$ and $\overline \eta$, which allows one to express $z$ in terms
of the latter two.  Indeed, the right-hand side (rhs) and the lhs of
\eref{equ8} should have the same scaling dimensions; therefore $ z =
d_\phi - d_{\overline \phi}$ in terms of the dimensions of the fields
in the time-domain.  Using now the expressions of the field anomalous
dimensions provided above, transforming into the dimensions in the
frequency domain, and replacing the Gaussian values in \eref{equZb}
and \eref{equZc} one concludes that
\begin{equation}\label{equ112} z = z_0 +
\frac{\overline{\eta}-\eta}{2}.
\end{equation}

\subsection{Perturbative expansion}
\label{Sec:perturbative}

\begin{figure}[!h]
\begin{center}
\begin{tabular}{cc}
\includegraphics[scale=1]{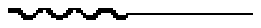}
& \includegraphics[scale=1]{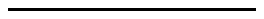} \\
$\langle\overline{\phi}_i(-\vec p, -\omega)\phi_j(\vec
p,\omega)\rangle$ & $\langle\phi_i(-\vec p, -\omega)\phi_j(\vec p,
\omega)\rangle$\\[2mm] (a) & (b)\\[4mm]
\includegraphics[scale=1]{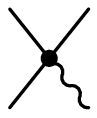}
& \includegraphics[scale=0.9]{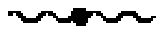}\\ $g$ &
$\Gamma_{i\omega}+\Gamma^*_{i\omega}$\\[2mm] (c) & (d)
\end{tabular}
\end{center}
\caption{Diagrammatic elements of the perturbation theory: (a)
response propagator, (b) correlation propagator, (c) interaction
vertex and (d) noise vertex. The straight parts of each line are
associated to fields $\phi$, whereas wiggled lines correspond to
$\overline\phi$ fields.}
\label{fig0}
\end{figure}

As we explained above, one does not expect any modification of
equal-time correlation functions, as they are determined by a static
theory with the effective Hamiltonian $\cal H$ in \eref{equ1}. Hence,
we focus on the dynamical exponent $z$, the corrections to which can
be obtained on the basis of the standard perturbative method
consisting in a combined expansion in the coupling constant $g$ and in
the deviation $\epsilon=4-D$ from the upper critical dimensionality of
the model \cite{Pa,Zi96,Ma76,LeBe91}. In performing such an expansion
one also takes advantage of the fact that $g$ will eventually be set
to its fixed-point value $g^*= {\mathcal O}(\epsilon)$.  We remind
here that the inverse temperature $\beta$ has been eliminated by a
suitable redefinition of the fields and the coupling constant $g$.  In
the following we concentrate on the one-particle irreducible vertex
functions~\cite{Zi96,LeBe91} with $n$ external $\phi$-lines and
$\overline{n}$ external $\overline{\phi}$-lines, denoted\footnote{Our
notation differs from the standard one, that is
$\Gamma^{n,\overline{n}}$ for the 1PI-vertex functions, in order to
avoid confusion with the noise kernel.} by \be
\mathcal{V}^{n,\overline{n}} = \mathcal{V}_0^{n,\overline{n}} +
\mathcal{V}_1^{n,\overline{n}} + \mathcal{V}_2^{n,\overline{n}} +
\cdots \ee The subscripts indicate the order in the perturbation
series. For example, $\mathcal{V}_2^{n,\overline{n}}$ includes all
terms proportional to $g^2$, $g\epsilon$ and $\epsilon^2$. The Feynman
rules of this perturbative expansion are those associated with 
the statistical weight $\rme^{-\mathcal{S}}$ in
\eref{eq:MSR} and they are the same as in the white noise
case~\cite{Jascsc88,Zi96}, the only difference being in the form of
the Gaussian response and correlation functions. In the diagrammatic
representation of the perturbation series we shall indicate the
relevant propagators and vertices as depicted in Fig.~\ref{fig0}. Note
that the noise vertex $\Gamma_{i\omega} + \Gamma^*_{i\omega}$ [see
Fig.~\ref{fig0}(d)] is diagonal in frequency space (i.e., it amounts
to a multiplication by an $\omega$-dependent factor) whereas it is
non-local in the time domain.
In addition, we point out the fact that in principle the correlation
function can be obtained in the frequency domain as a multiplication
of two response functions by the noise vertex, which corresponds to a
convolution in the time domain.

%


\subsubsection{Renormalization of the noise vertex.}
\label{sec:NV}
Our interest here is to know whether the correlated noise modifies the
critical behavior of the model. Within the Gaussian approximation $z$
is given by \eref{equZa}, where we assumed that a white-noise vertex
is generated under renormalization, a fact that yields 
two distinct fixed points $P$ and $P_{\rm w}$: the former is
characterized by the colored noise and is stable for $\alpha<\alpha_c
\equiv 1$, whereas the latter is characterized by the white noise, is
stable for $\alpha>\alpha_c$, and it reduces to the standard Model A
dynamics.  We shall show that, on the one hand, expanding around $P$
(with $\gamma_{\rm w} = 0$) renormalization indeed generates a
supplementary white noise vertex $\gamma_{\rm w}\neq 0$ and, on the
other hand, such a vertex becomes relevant at a $D$- and $N$-dependent
value $\alpha_c(D,N)$, where $\alpha_c(D,N)$ shows corrections to the
Gaussian cross-over occurring at $\alpha_c=1$ for $D<4$.

\begin{figure}[h]
  \begin{center}
  \includegraphics[width=0.32\textwidth]{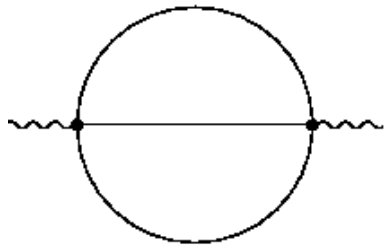}
  \end{center}
  \caption{Lowest-order perturbative contribution to the noise
vertex.}
  \label{fig1}
\end{figure}

The first correction to the noise vertex $\mathcal{V}_2^{0,2}$ is
given by the second-order diagram depicted in Fig.~\ref{fig1}
which can be conveniently written as the Fourier transform of its
expression in the time and space domain
\begin{eqnarray} \mathcal{V}_2^{0,2}(\vec q, \sigma) &=&
-\frac{g^2(N+2)}{18} \int\!\rmd^Dx\int_{- \infty}^{+\infty}\!\!\rmd t
\; \rme^{i\vec q\cdot \vec x - i\sigma t} \ C^3_0(\vec x,t) \
\nonumber\\\ &=& -\frac{g^2(N+2)}{9} \int_0^\infty \!\!\rmd
t\;\cos(\sigma t)\int\!\rmd^Dx\; {\rm e}^{i\vec q \cdot \vec x} \
C_0^3(\vec{x},t),
\label{equ11}
\end{eqnarray}
where the $N$-dependent prefactor accounts for the combinatorics of
the graph (see, e.g.,~\cite{Zi96}) and $C_0$ is the Gaussian
correlation function with $\gamma_{\rm w} = 0$. In the last line of
this equation we used the symmetry $C(\vec x,t) = C(\vec x,-t)$.
Since we are interested in the critical dynamics, we set $r$ to its
critical value $r_c = {\cal O}(g)$ (determined, e.g., by the value of
$r$ at which $C(\vec p=\vec 0,t=0)$ diverges~\cite{Zi96,Ma76,LeBe91}).
However, at the order $g^2$ we are presently interested in, one can
neglect the shift of the critical point and set $r=0$.
The leading behavior of the noise vertex is completely determined by
the small-$q$ and small-$\sigma$ asymptotics of
$\mathcal{V}_2^{0,2}$. We can set $q=0$ from the outset, while the
small-frequency limit $\sigma\rightarrow0$ has to be considered with
care since the tree-level noise vertex $\mathcal{V}_0^{0,2}(\sigma) =
2 {\rm Re}\Gamma_{i\sigma}(\gamma_{\rm w}=0) =
2\gamma|\sigma|^{\alpha-1}\sin(\pi\alpha/2)$ [see \eref{equN}]
diverges in this limit for $\alpha<1$. At the end we shall see that no
contribution to $\mathcal{V}_2^{0,2}$ proportional to
$|\sigma|^{\alpha-1}$ is actually generated. In what follows we only
take the limit $\sigma\rightarrow 0$ when it becomes manifestly
possible.
In this formulation, divergencies arise due to the singular behavior
of $C_0$ at small distances and times, $|\vec x|\to 0$ and
$t\rightarrow 0$. In order to regulize the theory, we introduce a
short-distance cut-off $\ell$, below which the description in the
continuum is no longer considered to be realistic. For example, in
lattice models, the cut-off $\ell$ is naturally identified with the
lattice spacing. Analogously, a short-time cut-off is introduced in
the convenient form $\ell^z$, which is motivated by the scaling
behavior discussed above. By using, c.f., the scaling form
\reff{equA4} of the Gaussian correlation $C_0$ influenced by the
colored noise (see App.~B), the second-order contribution to the
regularized vertex function takes the form
\begin{eqnarray} \mathcal{V}_2^{0,2}(\vec 0, \sigma; \ell) =
-\frac{g^2A_D(N+2)}{9} \int_{\ell^z}^\infty \rmd t \cos(\sigma
t)\int_\ell^\infty \rmd x\, x^{5-2D}g_{C_0}^3(\gamma x^2/t^\alpha),
\nonumber\\
\label{equ120}
\end{eqnarray} where $A_D= 2\pi^{D/2}/\Gamma_E(D/2)$ is the solid
angle in $D$ dimensions.

The Wilsonian renormalization scheme (see, e.g., \cite{Ma76}) amounts
to a resummation of the perturbation series which is performed
according to the following steps:

  (I) Effective vertex functions for the `slow' fluctuations are
determined by performing an integration (averaging) over 'fast'
fluctuations, within a spatial shell between $\ell$ and $b\ell$ and at
a temporal scale between $\ell^z$ and $(b\ell)^z$. As a result of this
integration the effective vertex functions --- and therefore the
coupling constants which characterize them --- acquire a dependance on
the scaling parameter $b>1$. To be more specific, consider the typical
integral which arises in loop calculations and which can be written in
the generic form
\begin{equation*} \mathcal I(\ell) = \int_{\ell^z}^\infty\rmd
t\int_\ell^\infty\rmd^D x \;\mathcal F(\vec{x},t) ,
\end{equation*} 
with some integrand $\mathcal F(\vec
x,t)$. The contribution of the integration over the fast fluctuations
is then equivalent to $\mathcal I(\ell) - \mathcal I(b\ell)$, an
expression which we shall use repeatedly below.  In the limit $b\to 1$
with $b>1$ one has $\mathcal I(\ell) - \mathcal I(b\ell) =
-\left[\partial \mathcal I(\ell) / \partial \ln\ell\right]\ln b +
\mathcal{O}(\ln^2b)$.

  (II) The effective vertex functions calculated in step (I) depend on
a new cut-off $b\ell$.  In order to recover the original cut-off
$\ell$ one rescales the coordinates and fields in the frequency and
momentum domain according to
\begin{eqnarray}\label{generalscaling}
  \begin{array}{rcl} \vec q &\mapsto& b^{-1}\vec q ,\\ \sigma
&\mapsto& b^{-z}\sigma ,\\ \phi_i &\mapsto& b^{D/2+1+z-\eta/2}\phi_i
,\\ \overline\phi_i &\mapsto&
b^{D/2+1+z-z_0-\overline\eta/2}\overline\phi_i .
  \end{array}
\end{eqnarray} The resulting expression is multiplied by $b^{-D-z}$
which accounts for the rescaling of the integration measure in the
Hamiltonian.

  (III) In order to study the evolution of the coupling constants
under the renormalization procedure it is convenient to consider the
case $b\to 1^+$ which corresponds to an infinitely thin integration
shell. In this case the evolution equations for the coupling constants
are coupled differential equations that depend upon $\alpha$ and the
anomalous dimensions introduced by the rescaling in step (II). The
anomalous dimensions are determined by requiring that all coupling
constants have a finite asymptotic value under the RG transformation
for $b\to\infty$.
 
Applying step (I) to the noise vertex function $\mathcal{V}_2^{0,2}$
we derive \eref{equ120} with respect to $\ln\ell$ and we multiply the
result by $\ln b$.  By defining [see \eref{equA7}]
\begin{equation}\label{eq:defE} u^2\mathcal{E}^{0,2}(\sigma;\gamma) =
\frac{\partial\mathcal{V}_2^{0,2}(\vec 0, \sigma;
\ell)}{\partial\ln\ell}
\end{equation} with \be\label{eq:defu} u = A_Dg/(2\pi)^D, \ee we find
that the effective noise vertex $ \mathcal{V}^{0,2}(\vec 0, \sigma\to
0; b\ell)$ for the slow fluctuations with short-time and -distance
cut-offs $b\ell$ and $(b\ell)^z$, respectively, is given by
\begin{eqnarray} \mathcal{V}^{0,2}(\vec 0, \sigma\to 0; b\ell)&=&
-(\Gamma_{i\sigma\to 0} + \Gamma^*_{i\sigma\to 0}) -
u^2\mathcal{E}^{0,2}(0;\gamma)\ln b \nonumber\\ &&
  \label{equ12} + {\mathcal O}(u^2\ln^2b, u^3).
  \end{eqnarray} For details on the calculation of
$\mathcal{E}^{0,2}(\sigma;\gamma)$ we refer to App.~\ref{Sec:A3}.

Clearly, the form of the effective noise vertex has changed, as the
term $\mathcal{E}^{0,2}(0;\gamma)$ generated by the non-Gaussian
fluctuations has the form of a white-noise contribution, whereas the
coefficient $\gamma$ of the colored noise is not modified up to this
order in perturbation theory. As a result, it is convenient to account
for the contribution of a white-noise vertex from the outset, by
replacing $\Gamma_{i\sigma}$ by $\Gamma_{i\sigma} + \gamma_{\rm w}$.
This implies that the Gaussian correlation function $C_0$ that
determines the loop correction still has a scaling form but with a
scaling function $g_{C_0}$ that is now a function of two variables,
see \eref{equA15}.  The correction $\mathcal{E}^{0,2}$ that is
generated depends on both $\gamma$ and $\gamma_{\rm w}$, we denote it
by $\mathcal{E}^{0,2}(0;\gamma,\gamma_{\rm w})$ and we explicitly
calculate it in \eref{equA7}.

The effective noise vertex depends on the cut-off $b\ell$. Following
step (II) of the renormalization procedure we rescale the effective
noise vertex as specified in \eref{generalscaling}. The coupling
strengths of the colored and the white noise $\gamma$ and $\gamma_{\rm
w}$ become running coupling constants $\gamma(b)$ and $\gamma_{\rm
w}(b)$ and in the limit $b\to 1$ they satisfy the set of coupled
differential equations \be
\label{equ16} \frac{\partial \gamma}{\partial\ln b} = \left[2-\alpha
z_0-\frac{\alpha}{2}(\overline{\eta}-\eta) - \eta\right]\gamma +
\mathcal{O}(\epsilon^3) \ee and \be
\label{equ17} \frac{\partial \gamma_{\rm w}}{\partial\ln b} = \left[2
- z_0 -\frac{\overline{\eta}+\eta}{2}\right]\gamma_{\rm w} +
\frac{z}{2}{u^*}^2\mathcal{E}^{0,2}(0;\gamma,\gamma_{\rm w}) +
\mathcal{O}(\epsilon^3), \ee valid at the critical point.
$u^*=\mathcal{O}(\epsilon)$ is the fixed point value of the coupling
constant, i.e., the value at which the effective coupling constant
$u(b)$ --- obtained by applying the procedure outlined here to the
$4$-point function --- flows for $b\to\infty$ and $D<4$. For $D>4$,
$u^*=0$ and the scenario within the Gaussian approximation presented
in Sec.~\ref{sec:Gaux} is not altered by the interaction. Accordingly
we focus below on the case $D<4$.
Two additional differential equations can be written by considering
how the coupling constant $u$ in $\mathcal{V}^{1,3}$ and the
coefficient of the term $\propto q^2$ in $\mathcal{V}^{1,1}(\vec
q,\ldots)$ are modified by the non-Gaussian fluctuations.  In
particular, the requirement of an effective $b$-independent
coefficient of $q^2$ fixes $\eta$ to its well-known static
value~\cite{Zi96} (see Sec.~\ref{sec:selfen-eta} for further details).

In order to determine the critical exponents we demand that the
amplitude of the noise vertex in the effective Hamiltonian be constant
as explained in step (III) of the renormalization procedure.
Neglecting for a while the contribution of the non-Gaussian
fluctuations to \eref{equ16} and \eref{equ17} (which amounts to
setting $u^*$ and the anomalous dimensions to zero), one can easily
solve them and recover the Gaussian picture which we anticipated in
Sec.~\ref{sec:Gaux}. Indeed, $\gamma(b)\sim b^{2-\alpha z_0}$ whereas
$\gamma_{\rm w}(b)\sim b^{2-z_0}$ as $b\rightarrow\infty$, which
implies $\gamma(b)/\gamma_{\rm w}(b)\sim
b^{(1-\alpha)z_0}$. Independently of the value of $z_0>0$, this ratio
tends to zero for $\alpha>1$.  The associated fixed point is
characterized by a finite $\gamma_{\rm w}(b\to \infty)$ with a
vanishing $\gamma(b\to\infty)$, i.e., the fixed point $P_{\rm w}$
introduced in Sec.~\ref{sec:Gaux}. In order for $\gamma_{\rm w}(b)$ to
stay finite, it is necessary to have $z_0 = z_0^{\rm (w)}=2$ in
\eref{equ17}, as expected from our previous discussion. On the
contrary, for $\alpha<1$, $\gamma(b)/\gamma_{\rm w}(b)\rightarrow
\infty$ for $b\rightarrow\infty$ and the associated fixed point has a
finite $\gamma(b\to\infty)$ and a vanishing $\gamma_{\rm
w}(b\to\infty)$, corresponding to the fixed point $P$ of
Sec.~\ref{sec:Gaux}. The former condition requires $z_0 = z_0^{\rm
(col)} = 2/\alpha$ in \eref{equ16}, consistently with the discussion
therein.

\begin{figure}[!h] \centering
 \includegraphics[width=0.65\textwidth]{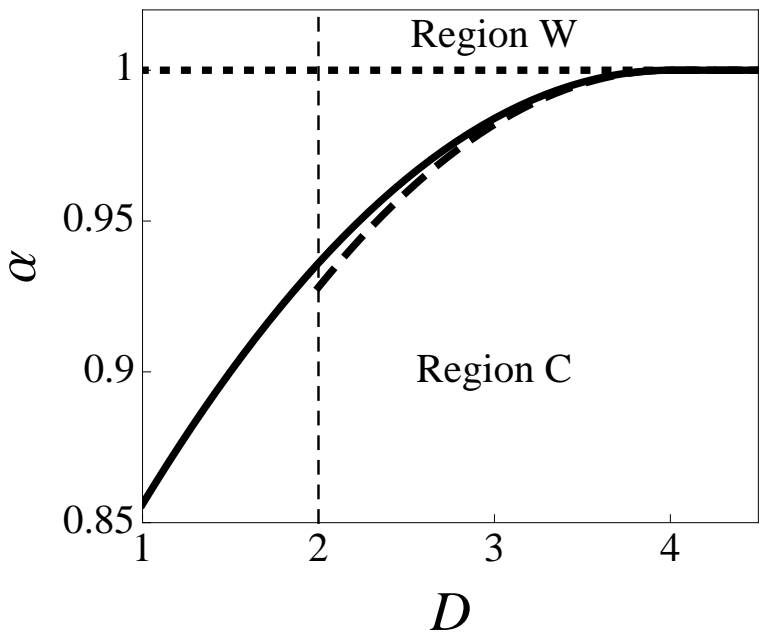}

  \caption{Boundary between the regions W and C of the
($\alpha,D$)-plane characterized, respectively, by white and colored
noise. The boundary curve $\alpha = \alpha_c(D,N)$ as a function of
the spatial dimensionality $D$ is reported here for $N=1$ (solid line,
Ising universality class), $4$ (dashed), and $\infty$ (dotted) where
the ${\cal O}((4-D)^3)$-correction in the corresponding perturbative
expression~\reff{equ18b} for $D<4$ has been neglected.  The vertical
dashed line indicates the lower critical dimensionality of the model
for $N>1$. The coefficient of the term ${\cal O}((4-D)^2)$ in
Eq.~\reff{equ18b} takes its maximum value for $N=4$ (dashed curve) and
then it decreases monotonically as a function of $N$, vanishing for
$N\to\infty$. For $D>4$, $\alpha_c$ takes the $D$-independent Gaussian
value $\alpha_{c,0} = 1$ (dotted line). Clearly, the dependence of the
boundary curve on the dimensionality $D$ is quantitatively rather
weak.}
  \label{fig3}
\end{figure}

Including now the effects of non-Gaussian fluctuations, the
colored-noise fixed point $P$ with $z_0=z_0^{\rm (col)} = 2/\alpha$
and $\gamma(b\rightarrow\infty)\neq 0$ is characterized by a value of
$\overline \eta$ such that the lhs of \eref{equ16} vanishes.  This
yields \be \overline\eta = \overline\eta^{\rm (col)} \equiv
\left(1-\frac{2}{\alpha}\right)\eta + {\cal O}(\epsilon^3) =
\left(1-\frac{2}{\alpha}\right)\frac{N+2}{2(N+8)^2}\epsilon^2 + {\cal
O}(\epsilon^3)\,.
\label{eq:etacol} \ee
We replaced $\eta$ by its static value given in ~\cite{Zi96,Ma76}
since, as we shall show in Sec.~\ref{sec:selfen-eta}, it is
$\alpha$-independent. The value $z^{\rm (col)}$ of $z$ at this fixed
point is determined via \eref{equ112} \be z = z^{\rm (col)} =
\frac{2-\eta}{\alpha} + {\cal O}(\epsilon^3) =
\frac{2}{\alpha}\left[1-\frac{N+2}{4(N+8)^2}\epsilon^2\right] +{\cal
O}(\epsilon^3).
\label{eq:zcol} \ee The fixed point $P$ is stable in the
$(\alpha,D)$-plane (region C in Fig.~\ref{fig3}) as long as the value
of $\gamma_{\rm w}(b)$ determined by \eref{equ17} at the fixed-point
$P$ with $\overline\eta = \overline\eta^{\rm (col)}$ and $z_0 =
z_0^{\rm (col)} = 2/\alpha$ stays finite for $b\rightarrow\infty$. A
crossover towards the fixed-point $P_{\rm w}$ in the
$(\alpha,D)$-plane (region W in Fig.~\ref{fig3}) controlled by the
white noise occurs as soon as $\gamma_{\rm
w}(b\rightarrow\infty)\rightarrow\infty$.  In this limit,
$\mathcal{E}^{0,2}(0;\gamma,\gamma_{\rm w}\to\infty) \simeq
\gamma_{\rm w}\mathcal{E}^{0,2}_{\rm w}$ independently of $\gamma$ [as
long as $\gamma(b)$ remains finite, see Eq.~(\ref{eq:E11}) for
details] where the constant $\mathcal{E}^{0,2}_{\rm w}$ is given in
\eref{equA9} and is such that \be {u^*}^2\mathcal{E}^{0,2}_{\rm w} =
\frac{N+2}{(N+8)^2}3\ln\frac{4}{3}\epsilon^2 + {\cal O}(\epsilon^3)\,.
\label{eq:e02asy} \ee Thus, the equation which determines the
evolution of $\gamma_{\rm w}$ at the fixed point $P$ becomes \be
\frac{\partial \gamma_{\rm w}}{\partial\ln b} = \left[2 - z_0^{\rm
(col)} -\frac{\overline\eta^{\rm (col)}+\eta}{2}+ \frac{z^{\rm
(col)}}{2}{u^*}^2\mathcal{E}^{0,2}_{\rm w}\right]\gamma_{\rm w} +
{\mathcal O}({u^*}^3), \ee and the crossover occurs as soon as the the
quantity in brackets changes sign.  The expression of the crossover
line is readily determined by taking into account the values of
$z_0^{\rm (col)}$, $\overline\eta^{\rm (col)}$, $z^{\rm (col)}$, and
$\mathcal{E}^{0,2}_{\rm w}$ reported in Eqs.~(\ref{eq:etacol}),
(\ref{eq:zcol}), and (\ref{eq:e02asy}): \be
\label{equ18b} \alpha_c= 1- \frac32 \ln\frac{4}{3} \
\frac{N+2}{(N+8)^2} \ \epsilon^2 + \mathcal{O}(\epsilon^3) .  \ee For
$\alpha > \alpha_c$ (region W in Fig.~\ref{fig3}), $\gamma_{\rm
w}(b\rightarrow\infty)\rightarrow\infty$ and the point $P$ is no
longer a fixed point of the rescaled effective action, as the
white-noise contribution becomes predominant. In order for it to
become constant and therefore to determine the fixed point $P_{\rm
w}$, $\overline\eta$ in \eref{equ17} should now take the value
$\overline\eta^{\rm (w)}$ such that $\partial\gamma_{\rm
w}/\partial\ln b=0$, with $z_0=z_0^{\rm (w)}=2$. Assuming that the
coefficient $\gamma(b)$ of the colored noise vanishes asymptotically
for $b\rightarrow\infty$, the lhs of \eref{equ17} becomes
$-(\overline{\eta}^{\rm (w)}+\eta)/2 +
(z/2){u^*}^2\mathcal{E}^{0,2}_{\rm w}$, where we used the fact that
$\mathcal{E}^{0,2}(0;\gamma=0,\gamma_{\rm w}) = \gamma_{\rm
w}\mathcal{E}^{0,2}_{\rm w}$. The condition that the rhs of the same
equation vanishes implies 
\be 
\fl\overline\eta = \overline\eta^{\rm
(w)} = -\eta + 2 {u^*}^2\mathcal{E}^{0,2}_{\rm w} + {\cal
O}(\epsilon^3) =
\frac{N+2}{(N+8)^2}\left[6\ln\frac{4}{3}-\frac{1}{2}\right]\epsilon^2
+ {\cal O}(\epsilon^3)
\label{eq:baretaw} \ee and from \eref{equ112}, \be z = z^{\rm (w)} = 2
+
\frac{N+2}{(N+8)^2}\left[3\ln\frac{4}{3}-\frac{1}{2}\right]\epsilon^2
+ {\cal O}(\epsilon^3)
\label{eq:zw} 
\ee 
in agreement with \cite{Bajawa76,Zi96}. In order to
verify the consistency of the assumption $\gamma(b)\rightarrow 0$ for
$b\rightarrow\infty$ under which \eref{eq:baretaw} has been derived,
one can specialize \eref{equ16} to the white-noise fixed point $P_{\rm
w}$, by using $\overline\eta^{\rm (w)}$, $z_0^{\rm (w)}=2$, and
$z^{\rm (w)}$ [see \eref{eq:zw}] as the values of $\overline\eta$,
$z_0$, and $z$. 
Accordingly, the term in
parenthesis in the rhs can be written as $-2(\alpha-\alpha_c) + {\cal
O}(\epsilon^3)$ and therefore $\gamma(b) \sim b^{-2(\alpha-\alpha_c)}$
indeed vanishes for $\alpha>\alpha_c$ as $b\rightarrow\infty$. This
also proves that the white-noise fixed point $P_{\rm w}$ is stable
against the perturbation of the colored noise as long as $\alpha >
\alpha_c$, a statement which complements the one presented above about
the stability of $P$.

Summarizing, \eref{equ18b} determines the line in the
$(\alpha,D)$-plane which separates region W from region C: in the
former, the white noise dominates and $z = z^{\rm (w)}$ (in agreement
with \cite{Bajawa76,Zi96}), whereas in the latter the colored noise
dominates and $z = z^{\rm (col)}$ is given by \eref{eq:zcol}.
Figure~\ref{fig3} illustrates this scenario for $N=1$, 4, $\infty$.

\subsubsection{Renormalization of the self-energy and FDT.}
\label{sec:selfen}
%
%
\begin{figure}[!h] \centering
  \includegraphics[width=0.32\textwidth]{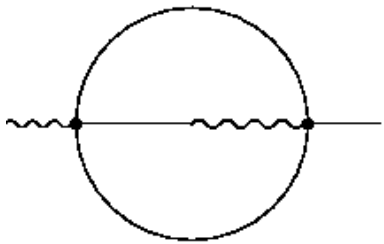}
  \caption{Second-order contribution to the self-energy.}
  \label{fig2}
\end{figure}
%
The fluctuation-dissipation theorem (FDT) expressed in \eref{equ8} is
a consequence of a symmetry of the action in
equilibrium~\cite{Jascsc88,Arbicu10} and it has to be preserved under
renormalization. Therefore, the noise vertex and the memory kernel
have to be related by the FDT even beyond the Gaussian approximation,
that we analyzed in Sec.~\ref{sec:Gaux}. Here we explicitly show that
this relation is still valid when non-Gaussian corrections up to
${\cal O}(\epsilon^2)$ (or, equivalently, ${\mathcal O}(g^2)$) are
accounted for. The first correction to the memory kernel comes from
the second-order self-energy contribution
\be
\label{equ19} \mathcal{V}_2^{1,1}(\vec q,\sigma) =
-\frac{g^2(N+2)}{6}\int_0^\infty \!\!\rmd t\int \rmd^Dx\;\rme^{
i\vec{q}\cdot\vec{x} - i\sigma t}\,C_0^2(\vec{x},t)R_0(\vec{x},t) \ee
represented in Fig.~\ref{fig2}. Note that $R_0$ is causal and
restricts the time integral to run over positive values only. The
expansion of this expression as a power series in $\sigma$ and
$\vec{q}$ allows one to identify the terms which contribute to the
renormalization of the different parameters of the Gaussian vertex
$\mathcal{V}_0^{1,1}(\vec q,\sigma) = i\sigma \Gamma_{i\sigma} + q^2 +
r$. The terms which are independent of both $\sigma$ and $\vec{q}$
contribute to the renormalization of the parameter $r$ (which is also
modified by an ${\cal O}(g)$ term not discussed here), the terms
proportional to $\sigma^0q^2$ contribute to the renormalization of the
fields and those proportional to $i\sigma^1 q^0$ to the
renormalization of the memory kernel $\Gamma$.  First of all we
observe that the FDT given in \eref{equ8} allows us to express $R_0$
in \eref{equ19} as $-\partial_tC_0$. An integration by parts yields
\begin{eqnarray} \mathcal{V}_2^{1,1}(\vec q,\sigma) =
-\frac{g^2(N+2)}{18}\int\rmd^Dx\;\rme^{ i\vec{q}\cdot\vec{x}}\left\{
C_0^3(\vec x,0) - i \sigma \int_0^\infty \!\!\rmd t\;\rme^{ - i\sigma
t}\,C_0^3(\vec{x},t) \right\}.  \nonumber\\
\label{eq:V11}
 \end{eqnarray}
Hence,
\begin{eqnarray}\label{eq:FDT2} \im \mathcal{V}_2^{1,1}(\vec q,\sigma)
&=& \sigma \frac{g^2(N+2)}{18} \ {\rm Re} \int\rmd^Dx\;\rme^{
i\vec{q}\cdot\vec{x}}\int_0^\infty \!\!\rmd t\;\rme^{ - i\sigma
t}\,C_0^3(\vec{x},t) \nonumber \\ &=& - \frac{\sigma}{2}
\mathcal{V}_2^{0,2}(\vec q,\sigma),
\end{eqnarray} where the last equality follows from a comparison with
\eref{equ11} and shows that the FDT in the frequency domain [see
\eref{eq:FDTomega}] is satisfied by the corrections ${\cal O}(g^2)$.
Note that the vertex $\mathcal{V}^{1,1}$ receives also a correction
$\mathcal{V}_1^{1,1}$ of ${\cal O}(g)$ given by a tadpole diagram
which, however, is a real constant and does not contribute to the
imaginary part. We conclude that up to and including the second order
in the coupling constant $ 2 \im \mathcal{V}^{1,1}(\vec{0},\sigma) = -
\sigma \mathcal{V}^{0,2}(\vec{0},\sigma)$.  (This proof can be readily
extended to the corresponding regularized vertex functions,
characterized by short- time and distance cut-offs.)

\subsubsection{Renormalization of the self-energy: the anomalous
exponent $\eta$.}
\label{sec:selfen-eta}

In the same spirit as before we can deduce the first correction to the
static exponent $\eta$. It is instructive to see why the dependence
upon $\alpha$ does not affect the final result, even though the
regularized expression of \eref{equ19} does via $C_0$ and $R_0$. In
order to single out the contribution of $\mathcal{V}^{1,1}_2$ to the
coefficient of $q^2$, one expands \eref{equ19} --- suitably
regularized as discussed above --- up to second order in $\vec q$,
finding
\begin{eqnarray} \fl \mathcal{V}_2^{1,1}(\vec q,\sigma=0;\ell) &=
\frac{1}{2}q^2 g^2\frac{N+2}{6}\frac{A_D}{D}\int_{\ell^z}^\infty \rmd
t\int_{\ell}^\infty \rmd x\;x^{D+1} \ C_0^2(\vec{x},t)R_0(\vec{x},t) +
\ldots,
\label{equ121}
\end{eqnarray} where the dots indicate all the terms which do not
contribute to the field renormalization, i.e., which are not
proportional to $\sigma^0q^2$. In \eref{equ121} we used the fact that,
for a generic function $f$, $\int {\rm d}^Dx \,x_ix_j \,f(|\vec{x}|) =
(\delta_{ij}/D)\int {\rm d}^Dx \, |\vec x|^2 f(|\vec{x}|)$, which is
valid also for the regularized integral.  As in the case of
\eref{eq:V11} one can take advantage of the FDT, \eref{equ8}, to
express the integrand in \eref{equ121} as a total derivative, which
can be integrated to yield
\begin{equation}\label{equ121b} \mathcal{V}_2^{1,1}(\vec q,0;\ell) =
q^2 g^2 \ \frac{A_{D}(N+2)}{36D}\int_{\ell}^\infty \rmd x\;x^{D+1} \
C_0^3(\vec{x},\ell^z) + \ldots
\end{equation} We note here that even though the (full) dynamic
correlation function $C$ (and therefore its Gaussian approximation
$C_0$) depends on the value of $\alpha$, the static correlation
function $C(\vec x,t=0)$ does not. This is explicitly shown for $C_0$
in \eref{equA3b}. While the limit $\ell\rightarrow 0$ of the rhs of
\eref{equ121b} cannot be explicitly taken due to the short-distance
singularity of the integrand, such a limit can be taken for the
correlation function, i.e., $C_0(\vec x,t=\ell^z) \simeq C_0(\vec
x,t=0)$ and therefore the expression of $\mathcal{V}_2^{1,1}(\vec
q,0;\ell\rightarrow 0)$ becomes --- as expected --- independent of
$\alpha$ at the leading relevant order in $\ell$.
Applying the same renormalization procedure as in Sec.~\ref{sec:NV},
\eref{equ121b} can be used to calculate the effective vertex
$\mathcal{V}_2^{1,1}(\vec q,0 ;b\ell)$ after having integrated out the
fast fluctuations. Similarly to \eref{eq:defE} one defines \be
q^2u^2\mathcal{E}^{1,1} + \ldots = - \frac{\partial
\mathcal{V}^{1,1}_2(\vec q,0 ;\ell) }{\partial \ln \ell} ,
\label{eq:defE11} \ee
For $b\rightarrow 1$, the resulting effective vertex is \be
\mathcal{V}_2^{1,1}(\vec q,0 ;b\ell) = q^2 + q^2u^2\mathcal{E}^{1,1}
\ln b + {\cal O}(u^2\ln^2 b,u^3) + \ldots
\label{equ21} \ee In order to recover the original cut-off $\ell$ we
rescale the fields and coupling constants according to
\eref{generalscaling} and we take $b\rightarrow 1$.  The part of
$\mathcal{V}_2^{1,1}(\vec q,0 ;\ell)$ that is proportional to
$\sigma^0q^2$ satisfies the evolution equation \be
\frac{\partial\mathcal{V}_2^{1,1}}{\partial\ln b} = -q^2\left[\eta -
{u^*}^2\mathcal{E}^{1,1}\right] + {\cal O}(u^2\ln b,u^3)+ \ldots.  \ee
By demanding that the amplitude of $\mathcal{V}_2^{1,1}$ be constant
and by using the numerical value of $\mathcal{E}^{1,1}$ calculated in
\eref{equA11} we find \be \eta = {u^*}^2\mathcal{E}^{1,1} =
\frac{N+2}{2(N+8)^2}\epsilon^2 + \mathcal{O}(\epsilon^3) , \ee i.e.,
$\eta$ has the same $\alpha$-independent value as in the static theory
confirming our expectations alluded to at the beginning of
Sec.~\ref{sec:eq}.


\section{Non-equilibrium dynamics}

\subsection{Preliminary remarks}
\label{sec:neqpr}

In this Section we investigate the non-equilibrium dynamics assuming
that the model is prepared in some initial condition at time $t=0$ and
that it is let relax subsequently at its critical point. This problem
has been studied in detail in the white-noise
case~\cite{Jascsc88,Jan92}.  The analysis reveals the emergence of an
interesting scaling behavior of two-time quantities, usually referred
to as \emph{aging} (see, in this context,
\cite{cg-05,Caga02,Cu-02}). More precisely, the relaxation is studied
via the field-theoretical action $\mathcal{S}$ in Eqs.~\reff{equ4} and
\reff{equ5} with $T=0$, supplemented by a suitable distribution
$\mathcal{P}_{IC}$ for the initial condition at time $t=T=0$. In
particular, a high-temperature disordered state is modeled by a
Gaussian distribution with zero mean:
\begin{eqnarray} 
\ln\mathcal{P}_{IC} &= - \int\rmd^Dx\;\frac{\tau_0}{2}\phi^2(\vec{x},t=0)
\end{eqnarray}
The parameter $\tau_0$ sets the inverse width of the initial
distribution. Within the Gaussian approximation, the field $\phi$ has
a scaling dimension $d_{\phi,0}$ given by \eref{equZa} in momentum and
frequency space, i.e., a dimension $d_{\phi,0}-z_0- D$ in the space
and time domain. Using this dimension for the initial field
$\phi(\vec{x},t=0)$ we find that $\tau_0\rightarrow b^2\tau_0$ under
rescaling. Consequently, the width of the initial distribution shrinks
to zero as $b\rightarrow \infty$, leading to a zero effective value of
the initial order parameter $\phi(\vec{x},t=0)=0$ and, therefore, to a
correlation function with Dirichlet boundary conditions at $t=0$.
In App. \ref{app:neq} [see \eref{equA15bis} and \eref{equA16}] we show
that the Laplace transforms of the Gaussian propagators are:
\begin{eqnarray}\label{nonE1} R_0(\vec{p};\lambda,\kappa) &=
\frac{1}{(\lambda+\kappa)(\lambda\Gamma_\lambda+p^2+r)},
\label{eq:neqR0}\\ C_0(\vec{p};\lambda,\kappa) &=
\frac{\Gamma_\lambda+\Gamma_\kappa}
{(\lambda+\kappa)(\lambda\Gamma_\lambda+p^2+r)(\kappa\Gamma_\kappa+p^2+r)}, \label{eq:neqC0}
\end{eqnarray} 
where the Laplace transformed noise is $\Gamma_\lambda
= \gamma\lambda^{\alpha-1} + \gamma_{\rm w}$ with $\lambda\in\mathcal
R_+$.
[In order to transform \eref{equ3} for $\alpha>1$ one should introduce
a short-time cut-off.  However, as we pointed out after \eref{equN},
this modification is not necessary as long as one is interested in the
leading long-time, near critical dynamic behavior of the system.
Accordingly, we shall use this form for $\Gamma_\lambda$ irrespective
of the value of $\alpha$.]
As in the equilibrium case, the propagators have a simple analytic
form in the time domain only for $\alpha=1$ or $\gamma_{\rm w} = 0$.
 It is easy to show that the response propagator is the same in and
out of equilibrium and, therefore, that it is time-translationally
invariant.

The correlation function $C_0$ can always be written as the sum of the
Gaussian equilibrium correlation $C^{\rm (e)}_0$ and the remaining
non-equilibrium contribution, which we denote by $C_0^{\rm (ne)}$ and
which will play an important role in fixing the genuinely
non-equilibrium properties of the relaxation, e.g., the
non-equilibrium exponent $\theta$ and the effective temperature.  The
Laplace transform $C^{\rm (e)}_0(\vec p,\lambda)$ of the equilibrium
correlation function $C_0^{\rm (e)}(\vec p,t)$ can be obtained from
Eq.~\reff{eq:C0-MF}: \be C^{(\rm e)}_0(\vec p,\lambda) =
\frac{1}{p^2+r}\frac{\Gamma_\lambda}{\lambda\Gamma_\lambda + p^2 + r}.
\ee The full non-equilibrium correlator \eref{eq:neqC0} can be
expressed as
\be\label{eq:C100} C_0(\vec p,\lambda,\kappa) = \frac{C^{\rm
(e)}_0(\vec p,\lambda)+C^{\rm (e)}_0(\vec
p,\kappa)}{\lambda+\kappa}-(p^2+r)C^{\rm (e)}_0(\vec p,\lambda)C^{\rm
(e)}_0(\vec p,\kappa) \ee which displays the fact that $C_0$ is the
sum of an equilibrium time-translationally invariant term and a
non-stationary term. Indeed, the Laplace transform $\mathcal
L[F](\lambda,\kappa) $ with respect to both $t$ and $t'$ of any
(translationally invariant) function $F(|t-t'|)$ is given by $\mathcal
L[F](\lambda,\kappa) = (F_\lambda+F_\kappa)/(\lambda+\kappa)$, which
is exactly the form of the first term in
\eref{eq:C100}~\cite{Po03}. Accordingly, we can identify the
non-equilibrium part $C_0^{\rm (ne)}$ of $C_0$ as $C_0^{\rm (ne)}(\vec
p,\lambda,\kappa) \equiv -(p^2+r)C^{\rm (e)}_0(\vec p,\lambda)C^{\rm
(e)}_0(\vec p,\kappa)$ which
translates, by virtue of \eref{eq:C0-MF}, into the
non-stationary expression ($t,t'>0$)
\be\label{eq:c200} 
C_0^{\rm (ne)}(\vec p;t,t') =
-\frac{1}{p^2+r}E_\alpha(-(p^2+r)t^\alpha/\gamma)E_\alpha(-(p^2+r){t'}^\alpha/\gamma)
, 
\ee
where we wrote the equilibrium Gaussian correlation function in terms
of the Mittag-Leffler function, anticipated in \eref{eq:C0-ML-ant} and
discussed in App.~\ref{Sec:Mittag}.

\subsection{General non-equilibrium renormalization group analysis}

\newcommand{\ii}{{{\rm in}}}

The addition of the initial condition to the action modifies the
scaling of the fields at the boundary $t=0$ compared to the one in the
`time bulk' $t>0$~\cite{Jascsc88}. In addition, to the bulk
renormalization a new initial time renormalization is required, which
gives rise to contributions `located' at the time surface (that is the
hyperplane determined by the condition $t=0$). These can be absorbed
by introducing a new anomalous dimension of the initial field
$\bar\phi_0(\vec{p})\equiv \overline{\phi}(\vec{p},0)$ (see
\cite{Didiei83} for an application to surface critical phenomena). The
general scaling of the initial fields in the time and momentum domain
reads [cf. Eqs.~\reff{equ110} and \reff{equ111}]
\begin{eqnarray}
  \label{equ:scal-initial-field}
  \begin{array}{rcl} \phi(b^{-1}\vec{p},t=0) &\mapsto&
b^{D/2+1-\eta/2}\;\phi(\vec{p},0),\\
\overline{\phi}(b^{-1}\vec{p},t=0) &\mapsto&
b^{D/2+1-z_0-\overline{\eta}/2-{\overline\eta}_{\ii}/2}\;\overline{\phi}(\vec{p},0),
  \end{array}
\end{eqnarray}
where ${\overline\eta}_\ii$ is a new exponent, with a Gaussian value
${\overline\eta}_{\ii,0}=0$. Note that the anomalous dimension of the
initial response field $\overline\phi(\vec p,0)$ is allowed to differ
by ${\overline\eta}_\ii$ from its bulk value. In~\cite{Jascsc88,cg-05}
one can find a careful analysis for the white-noise case where it is
explained why only the initial response field has
to be renormalized. Here, we make the same assumption and we check its
validity \emph{a posteriori}.  ${\overline\eta}_\ii$ is related to
the so-called initial-slip exponent $\theta$ [introduced at the end of
Sec.~\ref{sec:model}, see Eqs.~\reff{scalingR} and \reff{scalingC}]
by~\cite{Jascsc88} \be \theta = -\overline\eta_{\ii}/(2z) .
\label{eq:thetaeta} \ee
The analysis in~\cite{Jascsc88,cg-05} has to be slightly modified to
deal with colored noise.  Our starting point is the general leading
scaling behavior of the critical correlation functions
$\mathcal{G}^{n,\overline{n},\overline{n}_0}$ of $n$ bulk fields
$\phi$, $\overline{n}$ bulk response fields $\overline{\phi}$ and
$\overline{n}_0$ initial response fields $\overline{\phi}_0$,
evaluated at the set of points $\{\vec{p},t\}$ in momentum and time:
\begin{equation}\label{equG}
\mathcal{G}^{n,\overline{n},\overline{n}_0}(\{\vec{p},t\}) \simeq
b^{-\delta(n,\overline{n},\overline{n}_0)}\mathcal{G}^{n,\overline{n},\overline{n}_0}(\{b^{-1}\vec{p},b^zt\}),
\end{equation}
where $\delta(n,\overline{n},\overline{n}_0) = -D +
n(D/2+1-\eta/2)+\overline{n}(D/2+1-z_0-\overline{\eta}/2)+
\overline{n}_0(D/2+1-z_0-\overline{\eta}/2-\overline\eta_{\ii}/2)$.
[In writing \eref{equG} and the analogous relations presented below,
we always understand that the correlation functions on the lhs and rhs
are characterized by the different length cut-offs $b\ell$ and $\ell$,
respectively.]
This scaling behavior is a consequence of the scaling dimensions of
the fields $\phi$, $\overline\phi$ and $\overline\phi_0$ as functions
of time and momentum [compare to Eqs.~\reff{equ110}, \reff{equ111} and
\reff{equ:scal-initial-field}] and the dimension of the
$\delta$-function ensuring the total momentum conservation.
Note that all correlation functions with an initial field $\phi_0
\equiv \phi(\vec{p},0)$ vanish [see the discussion at the end of
Sec.~\ref{sec:neqpr}]. Specifically, the two point correlation and
response functions ($\mathcal{G}^{2,0,0}$ and $\mathcal{G}^{1,1,0}$,
respectively) scale as
\begin{eqnarray} C(\vec{p};t,t') &\simeq& b^{\eta-2}\;
C(\vec{p}/b;b^zt,b^zt') \label{eq:Cscale},\\ R(\vec{p};t,t') &\simeq&
b^{z_0-2+\eta/2+\overline{\eta}/2}\;
R(\vec{p}/b;b^zt,b^zt') \label{eq:Rscale}.
\end{eqnarray} By choosing $b=(t-t')^{-1/z}$ these scaling forms
become
\begin{eqnarray}\label{eq:scalingCR}
  \begin{array}{rcl} C(\vec{p};t,t') &\simeq& (t-t')^{(2-\eta)/z}\;
\tilde{F}_C((t-t')^{1/z}\vec{p},t'/t) ,\\ R(\vec{p};t,t') &\simeq&
(t-t')^{(2-z-\eta)/z}\; \tilde{F}_R((t-t')^{1/z}\vec{p},t'/t) .
  \end{array}
\end{eqnarray} In general the scaling functions $\tilde{F}_C$ and
$\tilde{F}_R$ are not expected to have a finite, non-vanishing value
for $t'\rightarrow 0$. In order to deduce their behavior for small
$t'$ we employ a short-distance expansion \cite{Zi96} of the fields
$\phi(\vec{p},t')$ and $\overline{\phi}(\vec{p},t')$ around
$t'=0$. 
However, these are not independent. Indeed, the
full correlation and linear response functions verify the equations
\cite{Jascsc88}
\begin{eqnarray}\label{eq:DysonC}
\begin{array}{rcl} C(\vec{p};t,t') &=& \int_0^t \rmd s \int_0^{s} \rmd
s' \;
R_0(\vec{p};t,s)\tilde\mathcal{V}^{1,1}(\vec{p},s,s')C_0(\vec{p};t',s')
\\ && + \int_0^{t'} \rmd s' \int_0^{s'}\rmd s \;
C_0(\vec{p};t,s)\tilde\mathcal{V}^{1,1}(\vec{p},s',s)R_0(\vec{p};t',s')
\\ && + \int_0^t \rmd s \int_0^{t'}\rmd s'\;R_0(\vec{p};t,s)
\tilde\mathcal{V}^{0,2}(\vec{p},s,s')R_0(\vec{p};t',s')
\end{array}
\end{eqnarray} and
\begin{eqnarray}\label{eq:DysonR} R(\vec{p};t,t') = \int_{t'}^t\rmd s
\int_{t'}^{s} \rmd s' \;R_0(\vec{p};t,s)
\tilde\mathcal{V}^{1,1}(\vec{p},s,s')R_0(\vec{p};s',t'),
\end{eqnarray}
where $\tilde\mathcal{V}^{n,\overline n}$ are the (not necessarily one
particle irreducible in contrast to the ones introduced in
Sec.~\ref{Sec:perturbative}) vertex functions with $n$ amputated
external field and $\overline n$ amputated external response field
legs, respectively. In writing these expressions we accounted for the
causality of $\tilde\mathcal{V}^{1,1}(\vec{p},s,s') \propto
\Theta(s-s')$.
After taking the derivative of \eref{eq:DysonC} with respect to $t'$
only the first term in the rhs survives in the limit $t'\to 0$ [note
that $R_0(\vec p;s,s)=0$~\cite{Arbicu10}]. By comparing the resulting
expression with the rhs of \eref{eq:DysonR} one notices that the
equations differ only by the last factor in their integrands,
$\partial_{t'}C_0(\vec p;t',s')$ and $R_0(\vec p;s',t')$,
respectively. We deduce that if a relation between the dimensions of
the time-derivative of the initial field and the initial response
field exists within the Gaussian approximation, it should be preserved
when non-Gaussian fluctuations are accounted for. Let us then examine
the propagators. We focus on region C where they satisfy the equation,
\begin{eqnarray}\label{eq:initialFDT}
  \partial_{t'} C_0(\vec{p};t,t'\rightarrow0)\simeq
{t'}^{\alpha-1}\int_0^t\rmd
s\,\Gamma(t-s)R_0(\vec{p};s,t'\rightarrow0)
\end{eqnarray} [proven in App.~\ref{app:neq}, see
\eref{eq:initialtimeFDT}]. In the early $t'$ limit we formally expand
the fields according to \be \phi(\vec{p},t'\to0)\sim
\varphi(t')\dot{\phi}_0(\vec{p})\quad \mbox{and} \quad
\overline{\phi}(\vec{p},t'\to0)\sim\overline{\varphi}(t')\overline{\phi}_0(\vec{p}).
\label{eq:ste} \ee $\phi$ is proportional to $\dot \phi_0(p)$ and
$\overline\phi$ is proportional to $\overline\phi_0(p)$ since the
former vanishes while the latter is allowed to be finite for $t'\to
0$.  We see from \eref{eq:initialFDT} that, under the rescaling
$t\rightarrow b^zt$, $s\rightarrow b^z s$ and $p\rightarrow p/b$
(leaving $t'$ unchanged), the scaling dimensions $d_{\dot{\phi}_0}$
and $d_{\overline{\phi}_0}$ of $\dot{\phi}_0$ and $\overline{\phi}_0$,
respectively, verify
\begin{equation}\label{eq:initialdim} d_{\dot{\phi}_0} = z(1-\alpha) +
d_{\overline{\phi}_0} .
\end{equation} For $\alpha=1$ this reduces to the relation found in
\cite{Jascsc88}.  The expansion of $\overline\phi$ in \eref{eq:ste}
can be used to calculate the correlation function
$\mathcal{G}^{1,1,0}(\vec p;t,t'\to 0)\sim \overline{\varphi}(t')
\mathcal{G}^{1,0,1}(\vec p;t)$ and by matching the scaling dimensions
of the lhs and rhs with the help of \eref{equG} we conclude that
$\overline{\varphi}(t') \sim {t'}^{-\theta}$ where $\theta$ is given
by \eref{eq:thetaeta}. Besides, the rescaling of $t'$ (keeping $t$ and
$s$ unchanged) implies $\varphi(t') \sim {t'}^{\alpha-\theta}$ if the
scaling dimensions of the lhs and rhs in \eref{eq:initialFDT} are to
match.  Hence, the small-$t'$ limit of the response function is
\begin{eqnarray} R(\vec{p};t,t'\rightarrow0) \sim \overline\varphi(t')
\langle\phi(-\vec{p},t)\overline{\phi}_0(\vec{p})\rangle \sim
{t'}^{-\theta} \mathcal{G}^{1,0,1}(\vec{p},t),
\end{eqnarray} where we introduced a short-hand notation for the
arguments of $\mathcal{G}^{1,0,1}$ in which we only write the
non-vanishing time $t$.  The scaling properties of
$\mathcal{G}^{1,0,1}(\vec{p},t)$ are given by \eref{equG}:
\begin{eqnarray}\label{eq:scalingG} \mathcal{G}^{1,0,1}(\vec{p},t)
&\simeq&
t^{-(\eta/2+\overline{\eta}/2+\overline\eta_{\ii}/2+z_0-2)/z}\;
\mathcal{G}^{1,0,1}(t^{1/z}\vec{p},1) \nonumber\\ &=& t^{(2-\eta
-z)/z+\theta}\; \mathcal{G}^{1,0,1}(t^{1/z}\vec{p},1)
\end{eqnarray} where we used the relation between anomalous and
dynamic exponents, \eref{equ112}, and the relation between $\theta$
and ${\overline\eta}_{\ii}$, \eref{eq:thetaeta}. Consequently, taking
Eqs.~(\ref{eq:scalingCR}), (\ref{eq:initialdim}) and
(\ref{eq:scalingG}) into account, we conclude that \be
R(\vec{p};t,t'\rightarrow0) \simeq t^{(2-z-\eta)/z}
\left(\frac{t}{t'}\right)^{\theta} F_R(t^{1/z}\vec{p},0).  \ee A
similar analysis of the scaling behavior of the correlation, taking
into account \eref{eq:initialdim}, yields \be
C(\vec{p};t,t'\rightarrow0) \simeq t^{(2-\eta)/z}
\left(\frac{t}{t'}\right)^{\theta-\alpha} F_C(t^{1/z}\vec{p},0).  \ee
These results are used to capture the singular behavior of the scaling
functions in \eref{eq:scalingCR} by writing:
\begin{eqnarray} R(\vec{p};t,t') \simeq (t-t')^{(2-z-\eta)/z}
\left(\frac{t}{t'}\right)^{\theta} F_R((t-t')^{1/z}\vec{p},t'/t),
 \label{eq:scalingformR} \\ C(\vec{p};t,t') \simeq (t-t')^{(2-\eta)/z}
\left(\frac{t}{t'}\right)^{\theta-\hat{\alpha}}F_C((t-t')^{1/z}\vec{p},t'/t),
  \label{eq:scalingformC}
\end{eqnarray} 
with
\begin{equation} \hat{\alpha} = \cases{ 1 \qquad\qquad \mbox{for} \;\;
\alpha \geq \alpha_c(D,N), \\ \alpha \qquad\qquad \mbox{for} \;\;
\alpha < \alpha_c(D,N), }
\end{equation} which encompass the white noise result $\hat\alpha =
1$~\cite{Jascsc88} for $\alpha \ge \alpha_c$.  The scaling functions
$F_C$ and $F_R$ are regular for $t'\rightarrow 0$ and depend on
$\alpha$. Moreover, in the RG sense they are universal functions up to
an overall amplitude and the normalization of their first argument.

The emergence of $\hat \alpha \neq 1$ for colored noise can be checked
within the Gaussian approximation by looking at the initial-slip
behavior of the propagators $R_0$ and $C_0$ with $\alpha<1$.  First of
all, note that $\theta$ takes the value $\theta_0=0$ within the
Gaussian theory, as one can infer by comparing the scaling form
\eref{eq:scalingformR} with the expression for the non-equilibrium
response $R_0$ at criticality, which coincides with the equilibrium
one in \eref{eq:R0-ML-ant} and is invariant under time
translations. Using this value $\theta_0$ of $\theta$ one has
$\displaystyle\lim_{\kappa\rightarrow\infty}\kappa
C_0(\vec{p};\lambda,\kappa)\sim \kappa^{-\alpha}$ and
$\displaystyle\lim_{\kappa\rightarrow\infty}\kappa
R_0(\vec{p};\lambda,\kappa)\sim \kappa^0$ from Eqs.~\reff{eq:neqR0}
and \reff{eq:neqC0}, respectively.

\subsubsection{The initial-slip exponent $\theta$.}

\begin{figure}[!h] \centering
  \includegraphics[width=0.15\textwidth]{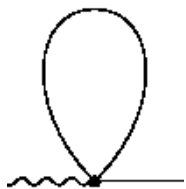}
  \caption{First order contribution to the non-equilibrium
self-energy.}
  \label{fig4}
\end{figure}

Out of equilibrium the first correction to the self energy leads to a
modification of the scaling of the initial response field.  The
response function up to first order in the perturbative expansion
reads, for zero external momentum,
\begin{eqnarray}\label{eq:Rnoneq} R(\vec{0};t,t';\ell) &=
R_0(\vec{0};t,t') + \int_{t'}^t\!\!\rmd
s\;R_0(\vec{0};t,s)B_{\ell^{-1}}(s)R_0(\vec{0};s,t').
\end{eqnarray} $B_{\ell^{-1}}(s)$ stands for the `tadpole' diagram
represented in Fig.~\ref{fig4}, which can be calculated by using
standard Feynman rules in the time domain
\cite{Masiro73,deDo76,Hoha77}, whereas $\ell^{-1}$ is the
large-momentum cut-off introduced in order to regularize the otherwise
divergent integral defining $B_{\ell^{-1}}(s)$:
\be B_{\ell^{-1}}(s) = -\frac{g(N+2)}{6}\int_{|\vec p|<\ell^{-1}}
\frac{\rmd^Dp}{(2\pi)^D} \ C_0(\vec p;s,s).
\label{eq:tadpole} \ee
 The renormalization of the initial response field is due
to the non-equilibrium part $C_0^{\rm (ne)}$ of $C_0$.  Indeed, the
equilibrium part $C_0^{\rm (e)}$ is characterized by time-translation
invariance and therefore it contributes with a time-independent
function of $\vec p$ to $C_0(\vec p;s,s)$ in \eref{eq:tadpole}. In
turn, such a function results in a time-independent contribution
$B_{\ell^{-1}}^{\rm (e)}$ to $B_{\ell^{-1}}$,
which can be thought of as due to a shift $r \mapsto r -
B_{\ell^{-1}}^{\rm (e)}$ of the mass $r$ in the expression of the
response functions $R_0$ appearing in the rhs of \eref{eq:Rnoneq},
i.e., as a mass renormalization.
[We recall that $R_0(\vec{p};t,t')$ actually depends on the two times
via $t-t'$.]
One can check that this term yields the correct first order correction
to the critical exponent $\nu$ which is the same as in the static
theory.

\begin{figure}[!t]
  \begin{center}
  \includegraphics[width=0.5\textwidth]{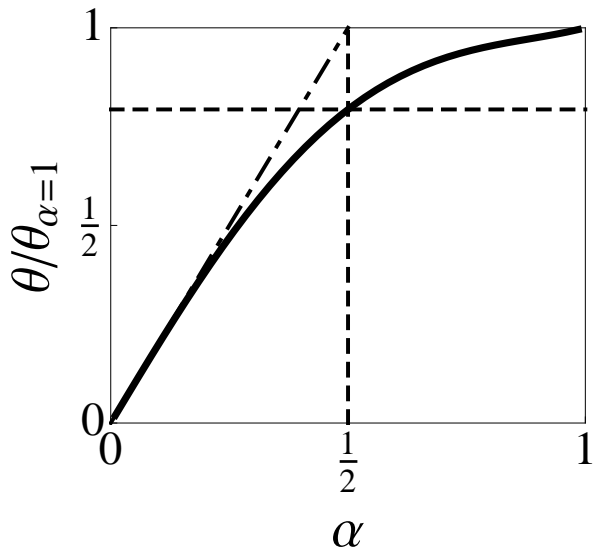}
  \end{center}
  \caption{Ratio $\theta/\theta_{\alpha=1}$ between the initial-slip
exponent $\theta$ in \eref{eq:finaltheta} and its white-noise value
$\theta_{\alpha=1}$, as a function of $\alpha$, within the relevant
range $0<\alpha \le 1$ at the first order in the
$\epsilon$-expansion. The dashed horizontal line indicates the value
$d(1/2)/\sqrt{\pi} = (\sqrt{2}/\pi) \ln(3+2\sqrt{2}) = 0.793\ldots$
corresponding to $\alpha=1/2$ (vertical dashed line). The dash-dotted
line points out the linear behavior $\theta/\theta_{\alpha=1} \simeq 2
\alpha$ expected for $\alpha\rightarrow 0$.}
  \label{fig:theta}
\end{figure}

In view of the renormalization procedure outlined in Sec.~\ref{sec:NV}
we need to calculate \be
\ell^{-1}\partial_{\ell^{-1}}B_{\ell^{-1}}(t)=-\frac{u(N+2)}{6}\ell^{-4}\
C_0^{\rm (ne)}(|\vec p|=\ell^{-1};t,t) \ee in the limit $\ell\to 0$,
for $r=0$ and $D=4$.
By using the asymptotic expansion of the generalized Mittag-Leffler
functions~\eref{eq:ML-asy} and their definition, \eref{def:genML}, one
finds
\be E_\alpha(x) = E_{\alpha,1}(x) = \cases{
(-x)^{-1}/\Gamma_E(1-\alpha) &\quad \mbox{for}\quad $x \rightarrow
-\infty$\\ E_\alpha(0)= 1 &\quad \mbox{for}\quad $|x| \ll 1$} \ee and
therefore, using \eref{eq:c200},
\begin{eqnarray} \fl
\label{eq:Rasympt} \ell^{-4}C_0^{\rm (ne)}(\ell^{-1};t,t)= - \ell^{-2}
E_\alpha^2(-t^\alpha/(\gamma\ell^2)) = \cases{
\mathcal{O}(\ell^2/t^{2\alpha}) &\mbox{for}\quad $\ell^{-2}
t^\alpha/\gamma \gg 1$,\\ \mathcal{O}(\ell^{-2}) &\mbox{for}\quad
$\ell^{-2} t^\alpha/\gamma \ll 1$.\\}
\end{eqnarray}
Accordingly, $\ell^{-1}\partial_{\ell^{-1}}B_{\ell^{-1}}(t)\to 0$ in
the limit $\ell\to 0$ for every fixed $t>0$. The physical
interpretation of this fact is that only the initial field is
renormalized by \eref{eq:Rnoneq}. Indeed the rhs of \eref{eq:Rasympt}
for finite $\ell$ provides an approximation of the delta distribution
restricted to $z\in {\mathbb R}^+$, usually denoted by $\delta_+(z)$:
\be (\gamma \ell^2)^{-1} E^2_\alpha(-z/(\gamma\ell^2)) \longrightarrow
\frac{d(\alpha)}{2} \delta_+(z) \quad\mbox{for}\quad \ell \rightarrow
0,
\label{eq:delta-p} \ee where the normalization constant $d(\alpha)$ is
given by \be d(\alpha) = 2 \int_0^\infty\!\! \rmd z\, E^2_\alpha(-z) ,
\ee and the additional factor $1/2$ on the rhs of \eref{eq:delta-p}
has been introduced for later convenience in order to have $d(1)=1$.
Taking advantage of the closed-form expressions of the Mittag-Leffler
function for $\alpha = 1$, $1/2$, and $0$, i.e., $E_1(-z) = \exp(-z)$,
$E_{1/2}(-z) = (2/\sqrt{\pi})\int_z^\infty\!\!\rmd t
\,\rme^{z^2-t^2}$, and $E_0(-z) = 1/(1+z)$~\cite{Hamasa09},
respectively, it is possible to calculate the corresponding values of
the $\alpha$-dependent constant $d(\alpha)$. One finds $d(1) = 1$,
$d(0)=2$ and, after some algebra, $d(1/2) = \sqrt{2/\pi}
\ln(3+2\sqrt{2}) = 1.406\ldots$.
Hence, \be \frac{\partial B_{\ell^{-1}}(t)}{\partial\ln\ell^{-1}}
\longrightarrow \frac{u\gamma(N+2)d(\alpha)}{12} \ \delta_+(t^\alpha)
\quad\mbox{for}\quad \ell \rightarrow 0.
 \label{eq:deltap} \ee
Using this expression and the one of the zero-momentum response
function $R_0(\vec 0;t,s) = (t-s)^{\alpha-1}/[\gamma
\Gamma_E(\alpha)]$ at criticality which follows from \eref{eq:R0-MF},
the derivative of the tadpole contribution to the rhs of
\eref{eq:Rnoneq} can be written as \be \fl
\frac{\partial}{\partial\ln\ell^{-1}} \int_{s}^t\!\!\rmd
s\;R_0(\vec{0};t,s)B_{\ell^{-1}}(s)R_0(\vec{0};s,t') = \delta_{t',0}
\frac{u(N+2)}{12}\frac{d(\alpha)}{\alpha\Gamma_E(\alpha)}
R_0(\vec{0};t,t')
\label{eq:derTP} \ee where $\delta_{t',0} = 1$ for $t'=0$ and $0$
otherwise, illustrating the fact that only the initial field is
renormalized. In deriving this last equation we used the fact that
$\delta_+(t^\alpha) = \delta_+(t)/(\alpha t^{\alpha-1})$.
Altogether, the effective response function with cut-off $b\ell$ reads
\begin{eqnarray} R(\vec{0};t,t';b\ell) = R_0(\vec 0;t,t')\left[1 +
\delta_{t',0}\frac{u(N+2)d(\alpha)}{12\alpha\Gamma_E(\alpha)}\ln
b\right].
\end{eqnarray}
In order to recover the original cut-off $\ell$ we make use of the
scaling relation \eref{equG} with $t'= 0$. By taking into account that
$\eta = \overline\eta = \mathcal O(\epsilon^2)$ we have
\begin{eqnarray} 
R(\vec{0};t,0;b\ell) &\simeq
b^{-2+z_0+\overline\eta_{\rm in}/2} R(\vec{0};b^zt,0;\ell) \\ &=
R_0(\vec 0;t,0)\ b^{\overline\eta_{\rm in}/2}\left[1 +
\frac{u(N+2)d(\alpha)}{12\alpha\Gamma_E(\alpha)}\ln b\right]\\ &=
R_0(\vec 0;t,0)\left[1 + \frac{\overline\eta_{\rm in}}{2}\ln b +
\frac{u(N+2)d(\alpha)}{12\alpha\Gamma_E(\alpha)}\ln b\right].
\end{eqnarray}
By requiring that the amplitude of the response function be constant
at the fixed point $u^*$ we obtain \be \overline\eta_{\ii} =
-\frac{(N+2)d(\alpha)}{(N+8)\alpha\Gamma_E(\alpha)}\epsilon + {\cal
O}(\epsilon^2) \ee whence we find the $\alpha$-dependent initial slip
exponent from \eref{eq:thetaeta}
\begin{equation} \theta = - \frac{\alpha}{4}\overline\eta_{\rm in}=
\frac{(N+2)d(\alpha)}{4(N+8)\Gamma_E(\alpha)}\epsilon + {\cal
O}(\epsilon^2).
  \label{eq:finaltheta}
\end{equation} In the white noise case $\alpha = 1$ we obtain $\theta
= (N+2)\epsilon/[4(N+8)] + {\cal O} (\epsilon^2)$ in agreement with
the first order result reported in~\cite{Jascsc88}. The dependence of
$\theta$ on $\alpha$ is shown in Fig.~\ref{fig:theta}. $\theta$
increases monotonically from $\theta=0$ at $\alpha=0$ to
$\theta=\theta(1)$ at $\alpha=1$ which is the cross-over value up to
$\mathcal{O}(\epsilon)$.
%
%
%
%

%
%
%

\subsubsection{Fluctuation-dissipation ratio and effective
temperature.}

A system which equilibrates after a certain finite relaxation time
satisfies the FDT. More generally, one defines the
\emph{fluctuation-dissipation ratio} (FDR) by
\begin{equation}\label{eq:FDR} X(\vec{p};t,t') =
\frac{\beta^{-1}R(\vec{p};t,t')}{\partial_{t'}C(\vec{p};t,t')},
\end{equation} where $\beta^{-1}$ is the temperature of the thermal
bath (set to $1$ in the previous analysis).  In glassy and weakly
driven macroscopic systems with slow dynamics --- small entropy
production limit --- this ratio approaches a constant on asymptotic
two-time regimes in which, moreover, it is independent of the
observable used to define the correlation and associated linear
response and admits the interpretation of an effective
temperature~\cite{Cukupe97,Cuku00}.  For systems with critical points,
the asymptotic value
\begin{equation}
X^\infty=\lim_{t'\to\infty}\lim_{t\to\infty}X(\vec{0},t,t')
\label{eq:Xinfty},
\end{equation} has been suggested to behave as a universal
property~\cite{GL00} and, moreover, as an \emph{effective
temperature},
\begin{equation} \beta^\infty = \beta X^\infty .
\end{equation} (Note, however, that beyond the Gaussian approximation
such a temperature depends upon the observable used to define it
\cite{CaGa04}.)  In equilibrium one has $X^\infty=1$ (which is just a
reformulation of the FDT) and $\beta^\infty=\beta$. Instead,
$X^\infty\neq 1$ is a signal of an asymptotic non-equilibrium dynamics
and therefore we shall focus on this quantity for the dynamics we are
presently interested in.

Within the Gaussian approximation discussed in Sec.~\ref{sec:Gaux} the
fluctuation-dissipation ratio $X$ can be easily calculated from the
expressions in Eqs.~\reff{eq:R0-ML-ant} and \reff{eq:C0-ML-ant} [see
also Eqs.~\reff{eq:R0-MF} and \reff{eq:C0-MF}] for the response and
correlation function, respectively, in terms of Mittag-Leffler
functions:
%
%
\begin{figure}[!t]
  \begin{center}
  \includegraphics[width=0.45\textwidth]{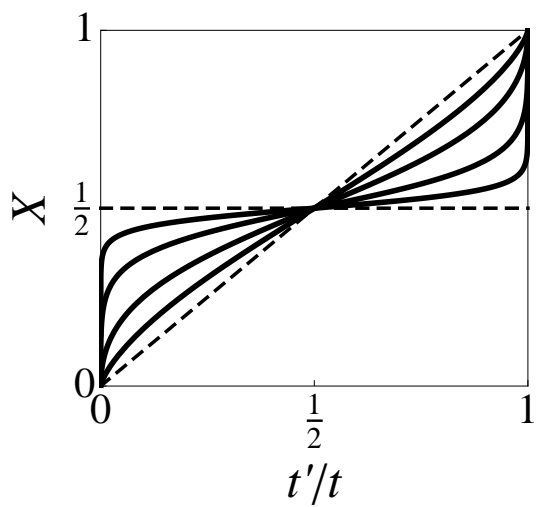}
  \end{center}
  \caption{Fluctuation-dissipation ratio for the global order
parameter (corresponding to $\vec{p}=0$) at criticality $r=0$ within
the Gaussian approximation, as a function of the ratio $0 \le t'/t \le
1$ for various values of $\alpha$. The straight horizontal and
diagonal dashed lines correspond to $\alpha=1$ and $\alpha=0$,
respectively.  The solid curves, instead, correspond to $\alpha=0.25$,
0.5, 0.75, and 0.9 upon moving away from the diagonal line.}
\label{fig:XGauss}
\end{figure}
%
\be X^{-1}(t,t') = 1 + \left(\frac{t}{t'}-1\right)^{1-\alpha}
\frac{E_\alpha(-At^\alpha/\gamma)E_{\alpha,\alpha}(-A{t'}^\alpha/\gamma)}{E_{\alpha,\alpha}(-A{(t-t')}^\alpha/\gamma)}
, \ee where we assumed $t>t'$ and $A \equiv p^2+r$. For $A\neq 0$
(e.g., far from the critical point $r=0$ or at criticality with
$\vec{p}\neq 0$) and long and well-separated times $t,t', t-t' \gg
(\gamma/A)^{1/\alpha}$, one can easily see from \eref{eq:ML-asy} that
$X^{-1} \rightarrow 1$, confirming the expectation that the system
equilibrates at long times, independently of the value of
$\alpha>0$. On the other hand, for the fluctuation of the homogeneous
mode $\vec{p}=0$ at criticality one has $A=0$ and the FDR takes the
simple form (originally derived in Ref.~\cite{Po03} for an anomalously
diffusing particle) \be X^{-1}_{\vec{p}=0, {\rm crit}} (t,t') = 1 +
\left(\frac{t}{t'}-1\right)^{1-\alpha} \ee which is a universal
scaling function of the dimensionless scaling variable $t'/t$,
reported in Fig.~\ref{fig:XGauss} for various values of $\alpha$. In
contrast to the white noise case $\alpha=1$, in the presence of
colored noise $0< \alpha <1$, $X^{-1}_{\vec{p}=0, {\rm crit}} (t,t')$
does actually depend on $t'/t$ and it interpolates continuously
between the quasi-equilibrium regime $t'\simeq t$, within which
$X_{\vec{p}=0, {\rm crit}} \simeq 1$, and the non-equilibrium regime
of well separated times $t' \ll t$, for which $X_{\vec{p}=0, {\rm
crit}} \simeq 0$ as it is generically observed in the case of
coarsening dynamics~\cite{Coliza,Cu11,Crri03}.

Beyond the Gaussian approximation, we can deduce an expression of the
two-time dependent FDR and its limiting values from the scaling forms
in \eref{eq:scalingformR} and \eref{eq:scalingformC}. First of all we
note that for $t\gg t'$ and $\vec{p}=0$ one has $\partial_{t'}C\simeq
t^{(2-\eta)/z+\theta-\hat{\alpha}}(\hat{\alpha}-\theta){t'}^{\hat{\alpha}-\theta-1}F_C(\vec{0},0)$
and $R\simeq t^{(2-\eta-z)/z}(t/t')^\theta F_R(\vec{0},0)$. We thus
obtain \be X^\infty =
\frac{F_R(\vec{0},0)}{(\hat{\alpha}-\theta)F_C(\vec{0},0)}
\displaystyle\lim_{t'\to\infty}\lim_{t\to\infty}\left(\frac{t}{t'}\right)^{\hat{\alpha}-1}.
\label{eq:Xinf} \ee
In the case $\hat\alpha=1$ of dominant white noise this expression
renders the well-known result $X^\infty =
F_R(\vec{0},0)/[(1-\theta)F_C(\vec{0},0)] |_{\alpha=1}\equiv
X^\infty_{\rm w}$ \cite{GL00,cg-05}, i.e., $X^\infty_{\rm w}=1/2$
within the Gaussian approximation~\cite{CKP94,CaGa02a,CaGa02b}. The
contribution of non-Gaussian fluctuations for $D<4$ and up to the
second order in the $\epsilon$-expansion have been calculated in
\cite{Caga02}, in rather good agreement with Monte Carlo simulation
(see Ref.~\cite{cg-05} for a summary). Instead, if the colored noise
dominates $\hat\alpha=\alpha < 1$ and therefore the long-time limit
$X^\infty$ of the FDR in \eref{eq:Xinf} vanishes, formally
corresponding to an infinite effective temperature as observed in
coarsening processes. Note that this result holds at all orders in
perturbation theory.
Therefore,
\begin{equation} X^\infty = \cases{ X^\infty_{\rm w} \qquad\qquad
\mbox{for} \;\; \alpha> \alpha_c(D,N) , \\ 0 \;\;\;\; \qquad\qquad
\mbox{for} \;\; \alpha< \alpha_c(D,N) , }
\label{eq:Xinf2}
\end{equation} where both values do not depend on the actual value of
$\alpha$ and therefore $X^\infty$ exhibits a discontinuity as a
function of $\alpha$ upon crossing the line $\alpha = \alpha_c(D,N)$.
Within the Gaussian approximation one can easily check the general
result \eref{eq:Xinf2} for $X^\infty$, on the basis of
Eqs.~\reff{nonE1} and \reff{eq:neqC0}. Indeed the behavior of the
correlation and response functions can be determined by taking
$\lim_{\lambda\to0}\lambda C_0(\vec{0};\lambda,\kappa)$ and
$\lim_{\lambda\to0}\lambda R_0(\vec{0};\lambda,\kappa)$, respectively,
for the propagators at zero momentum and at criticality. It is then
straightfoward to obtain $X^\infty_{0}=1/2$ within region W and
$X^\infty_{0}=\lim_{\lambda\to0}\Gamma_\kappa/\Gamma_\lambda = 0$
within region C, which confirms our general results.
Apparently, this result for $X^\infty_{0}$ contradicts the
corresponding one $X^\infty_{0}=1$ for a freely diffusing particle in
a super-Ohmic bath (corresponding to $\alpha>1$) found in \cite{Po03},
which our model reduces to within the Gaussian approximation.
However, within the field-theoretical approach discussed here, it
turns out that a super-Ohmic bath, responsible for a noise $\Gamma$
with $\alpha>1$ in \eref{equ3}, is eventually controlled by the
white-noise vertex and it is therefore unstable with respect to the
effects of the interaction, which effectively generates such a vertex
even though it was not present in the original coupling to the bath.
Therefore, the white-noise result $X^\infty_{0}=1/2$ does not only
apply to the cross-over line $\alpha_c(D,N)$ but it is valid within
the whole region W.
On the same footing, the results discussed here suggest that, at least
in higher spatial dimensions, adding interactions to a system which
displays superdiffusion (corresponding to $z<2$) results quite
generically in a sub-diffusive behavior ($z>2$) as expected in the
case of a diffusing particle (Gaussian approximation) with
interactions.


\section{Summary and outlook}

We studied the purely dissipative critical dynamics of a model with an
$N$-component order parameter in $D$ spatial dimensions, coupled to an
\emph{equilibrium} thermal bath which provides a colored thermal
noise.
We argued that the upper critical dimensionality of the model is
$D_c=4$ and we used the framework of the field-theoretical
$\epsilon$-expansion to account for the effects of non-Gaussian
fluctuations in $4-\epsilon$ spatial dimensions.

Within the Gaussian approximation --- valid for $D> D_c$ --- the
equilibrium dynamic exponent $z$ which controls the different scaling
of space and time takes the values \be z_0= \cases{ z_0^{(\rm col)} =
\ 2/\alpha &for $\alpha<1$,\\ z_0^{(\rm w )} \ = \ 2 &for $\alpha\geq
1$,\\}
  \label{eq:zconcl} \ee where $\alpha$ characterises the
\emph{algebraic} long-time decay of the two-time correlation function
of the noise, see \eref{equ3}. For $\alpha=1$ one recovers the
white-noise result $z^{(\rm w)}_0=2$.  The non-equilibrium `initial
slip exponent' $\theta$, instead, vanishes. Depending upon the value
of $\alpha$ the asymptotic long-time dynamics is effectively
equivalent to one driven by white noise (Ohmic bath) for $\alpha >
\alpha_c$, wheras the effect of the colored noise is relevant for
$\alpha<\alpha_c$. Within the Gaussian approximation $\alpha_c=1$, as
demonstrated by the change in behavior of $z_0$ given in
\eref{eq:zconcl}.

In dimensions $D<4$ the critical behavior is modified due to the
relevance of the interaction term and of the non-Gaussian
fluctuations. The value $\alpha_c$ which controls the cross-over
between the white-noise and the colored-noise dominated behaviors is
modified by $N$-dependent corrections of order $\epsilon^2$ and it
therefore separates the two corresponding regions in the parameter
space $(\alpha,D,N)$, named W and C in Fig.~\ref{fig3}, respectively.
The dynamical critical exponent $z$ is given by \be \fl z = \cases{
z^{\rm (col)} \equiv \frac{2}{\alpha} + \eta_\gamma =
\frac{2}{\alpha}\left[1-\frac{N+2}{4(N+8)^2}\epsilon^2\right] + {\cal
O}(\epsilon^3)&within region C, \\ z^{\rm (w)} \equiv 2+\eta_{\rm w} =
2 +
\frac{N+2}{(N+8)^2}\left[3\ln\frac{4}{3}-\frac{1}{2}\right]\epsilon^2
+ {\cal O}(\epsilon^3) &within region W.  \\} \ee
The $N$-dependent curve \reff{equ18b} which separates regions W and C
in the $(\alpha,D)$-plane is illustrated in Fig.~\ref{fig3} for $N=1$,
4, $\infty$.  Some comments are in order:
\begin{itemize}
\item[(i)] Upon decreasing $D$, the region W within which the Ohmic
result is recovered extends beyond the Gaussian value $\alpha_c=1$.
\item[(ii)] The correction to the Gaussian value $z_0$ is positive
within region W ($z_0=2$) and negative within region C
($z_0=2/\alpha$).
\item[(iii)] The exponent $z$ is a continuous function of $\epsilon$
and $\alpha$: At the transition line between regions W and C one has
$z^{\rm (w)} = z^{\rm (col)}$, as can be easily verified by using
\eref{equ18b}.
\item[(iv)] In the large-$N$ limit the $\epsilon^2$ correction
vanishes and the dynamic exponent $z$ and $\alpha_c$ take their
Gaussian values $z_0$ and $\alpha_c=1$, respectively.
\end{itemize}

For random initial conditions, i.e., with vanishing correlations and
average order parameter, we determined the general scaling forms of
the dynamic correlation functions.  Within region C, such scaling
forms differ from the ones valid in the presence of white noise only,
studied in Ref.~\cite{Jascsc88} and recovered within region W.  We
determined the corresponding initial-slip exponent $\theta$ up to
order $\mathcal{O}({\epsilon})$ in the presence of colored noise. It
is given by \be \theta =
\frac{(N+2)}{4(N+8)}{d(\alpha)}{\Gamma_E(\alpha)}\epsilon + {\cal
O}(\epsilon^2), \ee and the plot of the ratio between this value
$\theta$ and the reference $\theta_{\alpha=1}$ for the white noise is
reported in Fig.~\ref{fig:theta}.

In non-equilibrium conditions we also calculated the long-time limit
$X^\infty$ of the FDR for general $\alpha$ and $N$. The value of
$X^\infty$ in the presence of white noise is known analytically up to
$\mathcal{O}(\epsilon^2)$ \cite{Caga02} and numerically via Monte
Carlo simulations in various dimensions for models belonging to the
universality class of the $O(N)$ model with dissipative dynamics (see,
e.g., \cite{cg-05} for a review). We proved that this result is
recovered within region W.
Instead, if the colored noise is dominant [$\alpha < \alpha_c(D,N)$],
i.e., within region C, we showed that $X^\infty=0$. Therefore, the
associated effective temperature is infinite, analogously to what is
found in sub-critical coarsening~\cite{Coliza,Cu11}.  Our result for
$X^\infty$ within the Gaussian approximation is only in partial
agreement with the corresponding one derived in \cite{Po03} for an
anomalously diffusing particle --- i.e., of a fractional Brownian
motion --- which our model reduces to within such an
approximation. Indeed, in the presence of a super-Ohmic noise $\alpha
> \alpha_c=1$, one finds $X^\infty=1$~\cite{Po03} and super-diffusion
$z<2$ for the fractional Brownian motion, while we argue that
$X^\infty= X^\infty_0=1/2$ and normal diffusion $z=z_0^{\rm (w)}=2$ in
our field theoretical model. This is due to the fact that even in the
absence of a white-noise effective vertex in the original model,
non-Gaussian fluctuations (induced by the interactions) generate it
and turn it into the dominant one for $\alpha>\alpha_c\le 1$ such that
the white-noise result is recovered.

In conclusion, noises correlated in time may affect significantly the
equilibrium and non-equilibrium dynamical properties of systems close
to critical points.  In this respect it is important to note that the
distinction between super-Ohmic ($\alpha>1$) and sub-Ohmic
($\alpha<1$) thermal baths does not fully correspond to having
irrelevant (white) and relevant (colored) long-time correlations of
the noise, respectively.  Indeed, as shown in Fig.~\ref{fig3}, even a
weakly sub-Ohmic noise with $\alpha_c(D,N)<\alpha<1$ is actually
equivalent (in the RG sense) to an Ohmic (white) noise in the physical
dimensions $D=3$ and $D=2$ as far as the dynamical properties in the
long-time limit are concerned. In addition, in the presence of
interactions, a super-Ohmic bath does not result in a super-diffusive
behavior ($z<2$) but rather in the anomalous diffusion induced by the
equivalent white noise, in contrast to what happens for the free
fractional Brownian motion.

The field-theoretical predictions for the relaxational Markov critical
dynamics of systems belonging to the universality class considered
here have been put to the numerical test both via Monte Carlo
simulations and by solving the Langevin equations with a variety of
different methods (see, e.g.,~\cite{Ozit07} and references therein).
An instance of non-Markovian dynamics of the $\phi^4$-theory with a
noise exponentially correlated in time was investigated in
\cite{Sagagu98}. However, in this case one does not expect the
long-time dynamics of the system to be affected by the finite memory
of the noise. Dealing numerically with power-law correlated Gaussian
noise is a significantly harder problem which remains basically open
due to the difficulties in generating such kind of random process,
see, e.g., \cite{Zo09,Ba11} and references therein.

One of the virtues of the approach we have followed here is that it
can be easily applied to quantum critical
dynamics~\cite{Bocuga11b}. For instance, the thermal bath can be
modeled by a set of (quantum) harmonic oscillators coupled to all
degrees of freedom of the system. Within the Schwinger-Keldysh
formalism it is possible to derive a path-integral representation of
the non-equilibrium dynamics~\cite{We08,Ka05}. Integrating out the
oscillator variables one obtains an action similar to the one
considered here \cite{We08,Ka05,Grscin88}. The main difference with
the classical case is that even Ohmic dissipation leads to retarded
interactions. The present work is expected to provide at least a
partial and preliminary insight into the more difficult problem of the
analysis of quantum critical equilibrium and non-equilibrium
dissipative dynamics~\cite{Bocuga11b}.

Among other possible extensions of the present work, we mention the
problem of understanding the effects of colored noise on sub-critical
coarsening. The dynamic scaling hypothesis states that the late-stage
phase ordering kinetics is governed by a length scale $L(t)$ that, in
models with no quenched disorder, typically grows in time as a
power-law $L(t)\simeq \lambda(T) t^{1/z_d}$. The dynamic exponent
$z_{d}$ (generically different for the dynamic exponent $z$ at
criticality) depends upon the kind of order parameter and the
conservation laws~\cite{Br02} while the prefactor $\lambda(T)$
typically depends only weakly upon temperature $T$, is non-universal,
and it vanishes upon approaching a critical point. (The matching with
the critical growth is explained in~\cite{Siarbrcu07}.)  In presence
of colored noise this growth law might be modified, even though one
usually expects thermal fluctuations not to affect the domain growth
\cite{Br02}. We shall address this issue in a future study.

\ack AG is supported by MIUR within the program `Incentivazione alla
mobilit\`a di studiosi stranieri e italiani residenti all'estero' and
by CNRS (France).  This work was financially supported by
ANR-BLAN-0346 (FAMOUS).  LFC and AG thank the ICTP and LPTHE,
respectively, for hospitality during the preparation of this work. AG
thanks A. Pelissetto for valuable discussions.

\appendix
\begin{appendices}

\section{Fourier and Laplace conventions}

Within the present study we define the Fourier
transform and its inverse via
\begin{equation}\label{equA100} \hat{F}(\omega) =
\int_{-\infty}^\infty \rmd t\;\rme^{-i\omega t} \ F(t) ,
\qquad\mbox{and}\qquad
\label{equA101} F(t) =
\int_{-\infty}^\infty\frac{\rmd\omega}{2\pi}\;\rme^{i\omega t} \
\hat{F}(\omega) .
\end{equation} Instead, for every $\lambda>0$ the Laplace
transform is defined as
\begin{equation}\label{equA102} \hat F_\lambda = \int_{0}^\infty \rmd
t\;\rme^{-\lambda t}\ F(t) .
\end{equation} In the main text we shall drop the hats, whenever this
does not generate confusion. 


\section{The equilibrium propagators}

\subsection{Scaling in real time}
\label{sec:A2}

For $\alpha=1$ (white noise) the equilibrium propagators have a simple
analytic form in the time domain~\cite{Bajawa76,Jascsc88,Arbicu10}.
They can be calculated by applying an inverse Fourier transform to
Eqs.~(\ref{equR}) and (\ref{equC}):
\begin{equation}\label{equA1}
R_0(\vec{p},t)=\Theta(t)\rme^{-(p^2+r)t/\gamma_{\rm w}}\nonumber
\end{equation}
\begin{equation}\label{equA2}
C_0(\vec{p},t)=\frac{1}{p^2+r}\rme^{-(p^2+r)|t|/\gamma_{\rm w}} .
\end{equation}

For general $\alpha$ dimensional analysis suggests that the critical
($r=0$) Gaussian correlation $C_0$ with $\gamma_{\rm w} = 0$ should
scale as
\begin{equation}\label{equA3} C_0(\vec{p},t)= p^{-2}\
f_{C_0}(p^2|t|^\alpha/\gamma) .
\end{equation} Using Eqs.~(\ref{equN}) and (\ref{equC}) the equal-time
correlator is given by
\begin{eqnarray}\label{equA3b} C_0(\vec{p},t=0) &=
\int\frac{\rmd\omega}{2\pi}\ C_0(\vec{p},\omega) \nonumber\\
&=\int\frac{\rmd\omega}{2\pi} \
\frac{2\gamma\sin(\pi\alpha/2)|\omega|^{\alpha-1}}{\gamma^2|\omega|^{2\alpha}
+ 2\gamma(p^2+r)|\omega|^\alpha\cos(\pi\alpha/2)+(p^2+r)^2}
\nonumber\\ &=\frac{1}{p^2+r} .
\end{eqnarray} Hence, we infer that $f_{C_0}(0)=1$. Naturally, we have
$f_{C_0}(\infty)=0$ since correlations have to vanish in the long-time
limit. Applying a Fourier transform to Eq.~(\ref{equA3}) it is easy to
show that at criticality ($r=0$)
\begin{equation}\label{equA4} C_0(\vec{x},t)=\frac{1}{|x|^{D-2}}\
g_{C_0}(\gamma x^2/|t|^\alpha) ,
\end{equation} where the function $g_{C_0}$ reaches the asymptotic
value $g_{C_0}(\infty) = \Gamma_E(D/2-1)/(4\pi^{D/2})$.  In order to
deduce the leading behavior for $g_{C_0}(u)$ when $u\rightarrow 0$ we
start from the explicit expression of the noise kernel
$\Gamma_{i\omega}$ given in \eref{equN}. After some algebra we obtain
\begin{eqnarray} g_{C_0}(u) =
\int\frac{\rmd^Dp}{(2\pi)^D}\frac{\rmd\omega}{2\pi}\,
\frac{2u\sin(\pi\alpha/2)|\omega|^{\alpha-1}\rme^{i\omega+i\vec{p}\cdot\hat{z}}}{u^2|\omega|^{2\alpha}+2up^2|\omega|^\alpha\cos(\pi\alpha/2)+p^4}
,
\end{eqnarray} where $u=\gamma x^2/t^\alpha$ and $\hat{z}$ is an
arbitrary unit vector.  For $\alpha\leq 1$ we neglect the
contributions of $\mathcal{O}(u^2)$ in the denominator and we obtain
\begin{eqnarray} g_{C_0}(u\rightarrow 0) =
2u\int\frac{\rmd\omega}{2\pi}\,|\omega|^{\alpha-1}\rme^{i\omega}
\int\frac{\rmd^Dp}{(2\pi)^D}\,\frac{\sin(\pi\alpha/2)
\rme^{i\vec{p}\cdot\hat{z}}}{(p^2+u|\omega|^\alpha\cos(\pi\alpha/2))^2}.
\end{eqnarray} The integral over $\vec{p}$ is of
$\mathcal{O}(u\ln\left[u|\omega|^\alpha\cos{\pi\alpha/2}\right])$ for
$D=4$ and the resulting integral converges for $\alpha<1$;
consequently,
\begin{equation}\label{equA4b} g_{C_0}(u\rightarrow
0)\sim\mathcal{O}(u\ln u).
\end{equation} 
By using FDT we derive
\begin{equation}\label{equA5} 
R_0(\vec{x},t) =
\frac{\alpha\gamma}{x^{D-4}t^{\alpha+1}}\ g_{C_0}'(\gamma
x^2/t^\alpha)\Theta(t) .
\end{equation} In the white-noise case, the scaling function $g_{C_0}$
has the simple form
\begin{equation}\label{equA6} 
g_{C_0}(u) =
\frac{\Gamma_E(D/2-1)}{4\pi^{D/2}} \left[ 1- \frac{\Gamma_E\left(
\frac{D}{2}-1, \frac{u}{4}\right)}{\Gamma_E\left(\frac{D}{2}-1\right)}
+ \mathcal{O}(\epsilon) \right] ,
\end{equation} 
with $\Gamma_E(s,x) = \int_x^\infty \rmd y \ y^{s-1}
\rme^{-t}$, whence we deduce for $\alpha=1$ and $D=4$
\begin{equation}
\label{equA4c} 
g_{C_0}(u\rightarrow 0) = \mathcal{O}(u) .
\end{equation}

For generic $\gamma$ and $\gamma_{\rm w}$ the scaling function
$g_{C_0}$ is no longer a function of one variable. It is easy to show
that
\begin{equation}
\label{equA15} C_0(\vec{x},t)=\frac{1}{|x|^{D-2}} \ g_{C_0}(\gamma
|x|^2/|t|^\alpha,\gamma_{\rm w} |x|^2/|t|).
\end{equation} Moreover, by using a similar argument as above one has
for $D=4$
\begin{equation}
\label{equA16} \lim_{t\to\infty}g_{C_0}(u,v) = \mathcal{O}((u+v)\ln
u),
\end{equation} where $u = \gamma |x|^2/t^\alpha$ and $v = \gamma_{\rm
w} |x|^2/t$ vanish with $u/v$ finite. In the opposite short-time limit
in which $u$ and $v$ diverge with $u/v$ finite,
\begin{equation}
\label{equA166} \lim_{t\to0}g_{C_0}(u,v) =
\Gamma_E(D/2-1)/(4\pi^{D/2})
\end{equation} as for the purely colored problem.

The equilibrium propagators can be written in terms of the generalized
Mittag-Leffler functions $E_{a,b}(z)$, as discussed in
App.~\ref{Sec:Mittag}.


\subsection{Generalized Mittag-Leffler functions}
\label{Sec:Mittag}

The Laplace transform of $R_0(\vec{p},t)$ is given by \be
R_0(\vec{p},\lambda) = \frac{1}{\lambda\Gamma_\lambda + A} \ee where
we defined $A \equiv p^2+r$ and, in the case of colored noise,
$\Gamma_\lambda = \gamma \lambda^{\alpha-1}$.  We formally expand this
expression for small $A$: \be R_0 (\vec{p},\lambda) =\frac{1}{\gamma
\lambda^\alpha} \frac{1}{1+A(\gamma\lambda^\alpha)^{-1}} =
\frac{1}{\gamma \lambda^\alpha} \sum_{k=0}^\infty
\frac{(-A/\gamma)^k}{\lambda^{\alpha k}}
\label{eq:LplTrR0} \ee where the terms of the form $1/\lambda^\beta$
(with ${\mbox Re}\, \beta > 0$) are recognized as the Laplace
transform of $\Theta(t) t^{\beta-1}/\Gamma_E(\beta)$, so that
\reff{eq:LplTrR0} is identified as the Laplace transform of
\begin{eqnarray} R_0(\vec{p},t) &=& \Theta(t) \frac{1}{\gamma}
\sum_{k=0}^\infty (-A/\gamma)^k \frac{t^{\alpha
k+\alpha-1}}{\Gamma_E(\alpha k + \alpha)} \nonumber\\ &=& \Theta(t)
\frac{t^{\alpha-1}}{\gamma} E_{\alpha,\alpha}(-At^\alpha/\gamma) ,
\label{eq:R0-MF}
\end{eqnarray} where we have introduced the generalized Mittag-Leffler
function \be E_{\alpha,\beta}(z) \equiv \sum_{k=0}^\infty
\frac{z^k}{\Gamma_E(\alpha k +\beta)} \quad \mbox{with} \quad
\alpha,\beta,z \in {\mathbb C}, \ \mbox{Re}\{\alpha,\beta\} >0.
\label{def:genML} \ee Note that this function reduces to an
exponential for $\alpha =\beta =1$: $E_{1,1}(z) = \rme^z$, whereas for
$z\in {\mathbb R}$~\cite{Hamasa09}, \be E_{\alpha,\beta}(z\rightarrow
- \infty) = - \sum_{k=1}^{k^*}\frac{1}{\Gamma_E(\beta-\alpha k)}
\frac{1}{z^k}+ {\mathcal O}(z^{-(k^*+1)}) .
\label{eq:ML-asy} \ee

The corresponding expression for the equilibrium Gaussian correlation
function can be obtained from the FDT~\reff{equ8}. Indeed, after
integration \eref{equ8} takes the form \be C_0(\vec{p},t) =
C_0(\vec{p},t=0) - \int_0^{|t|}\!\!\rmd s \, R_0(\vec{p},s) \ee where
we used the fact that, in equilibrium, $C(\vec{x},t) =
C(\vec{x},-t)$. Taking into account \reff{equA3b} and the first line
of \reff{eq:R0-MF} one readily finds 
\be 
C_0(\vec{p},t) = \frac{1}{A}
E_{\alpha}(-A|t|^\alpha/\gamma)
\label{eq:C0-MF} 
\ee 
where $E_\alpha(z) \equiv E_{\alpha,1}(z)$ is the
Mittag-Leffler function.

The correlation function $C_0(\vec{p},t)$ in \eref{eq:C0-MF} can also
be expressed as the inverse Fourier transform of $C_0(\vec{p},\omega)$
reported in Eq.~\reff{equC} (see also Eq.~\reff{equN}). After some
suitable change of variables one finds the following scaling form \be
C_0(\vec{p},t) = \frac{1}{A} f_{C_0}(A|t|^\alpha/\gamma) , \ee where
\begin{eqnarray} f_{C_0}(u) &\equiv& \frac{2}{\pi} \int_0^\infty
\!\!\rmd v \cos(u^{1/\alpha} v)\frac{v^{\alpha-1}
\sin(\pi\alpha/2)}{v^{2\alpha} + 2 v^\alpha \cos(\pi\alpha/2) + 1}
\nonumber\\ & = & \frac{\sin(\pi\alpha/2)}{\pi \alpha/2} \int_0^\infty
\!\!\rmd v \frac{\cos(u^{1/\alpha} v^{1/\alpha})}{v^2 + 2 v
\cos(\pi\alpha/2) + 1}
\end{eqnarray} is the explicit expression for the scaling function
introduced in \eref{equA3}.


\section{Calculation of $\mathcal{E}_{\rm w}^{0,2}$ and
$\mathcal{E}^{1,1}$}\label{Sec:A3}

Starting from \eref{equ120} we have for generic $\gamma$ and
$\gamma_{\rm w}$
\begin{eqnarray}\label{equA7} &&
u^2\mathcal{E}^{0,2}(\sigma;\gamma,\gamma_{\rm w}) = \ell \
\frac{g^2A_D(N+2)}{9} \nonumber\\ && \qquad\;\;\; \times \left\{
z\ell^{z-1} \cos(\sigma\ell^z) \int_\ell^\infty \rmd x\; x^{5-2D} \
g_{C_0}^3 \left(\frac{\gamma x^2}{\ell^{\alpha z}},\frac{\gamma_{\rm
w} x^2}{\ell^{z}}\right) \right.  \nonumber\\ && \qquad\qquad\;\;\;
\left.  + \int_{\ell^z}^\infty\rmd t\,\cos(\sigma t) \ \ell^{5-2D} \
g_{C_0}^3 \left(\frac{\gamma \ell^2}{t^\alpha},\frac{\gamma_{\rm w}
\ell^2}{t}\right) \right\}.
\end{eqnarray}
The result of the integral in the first term in curly brackets is an
analytic function of $\sigma$ that admits a Taylor expansion in powers
of $\sigma^2$, i.e.,
\begin{equation} c_0 + c_2 \sigma^2 + c_4 \sigma^4 + \dots
\end{equation} with coefficients that, in principle, depend separately
on $\gamma$, $\gamma_{\rm w}$ and $\ell$. The integral in the second term in curly
brackets yields, instead, a non-analytic function of $\sigma$ that we
can still express as a series: \be d_0 + d_2 \sigma^2 + \dots +
d_{3\alpha-1} \sigma^{3\alpha-1} + \dots \ee where the term $\propto
\sigma^{3\alpha-1}$ is due to the leading behavior of $g_{C_0}^3$ for
$t\to+\infty$ [see \eref{equA4b}] which has to be subtracted for
$\alpha<1/3$ in order to make the integral convergent at large $t$.
If $3\alpha-1>0$ the limit $\sigma\to 0$ can be safely taken and the
white-noise vertex is renormalized by $c_0+d_0$.  If, on the contrary,
$3\alpha-1<0$ the contribution proportional to $\sigma^{3\alpha-1}$ is
anyhow negligible (for $\alpha>0$) with respect to the term
$\gamma\sigma^{\alpha-1}$ which is already present in the tree-level
vertex. Therefore, there is no renormalization of the colored-noise
vertex and we can focus on the limit $\gamma_{\rm w}\gg \gamma$, i.e.,
on the correction to the white-noise vertex only.
Since we calculate evolution equations up to order $\epsilon^2$ we
simply need to evaluate (\ref{equA7}) in $D=4$. We obtain
\begin{eqnarray}\label{equA7b} &&
u^2\mathcal{E}^{0,2}(0;\gamma,\gamma_{\rm w}) = \ell \
\frac{g^2A_D(N+2)}{9} \nonumber\\ && \qquad\;\;\; \times \left\{
z\ell^{z-1} \int_\ell^\infty \rmd x\; x^{-3} \
g_{C_0}^3\left(\frac{\gamma x^2}{\ell^{\alpha z}},\frac{\gamma_{\rm w}
x^2}{\ell^{z}}\right) \right.  \nonumber\\ &&
\qquad\qquad\qquad\qquad\;\;\;\; \left.  + \int_{\ell^z}^\infty\rmd
t\, \ell^{-3} \ g_{C_0}^3\left(\frac{\gamma
\ell^2}{t^\alpha},\frac{\gamma_{\rm w} \ell^2}{t}\right) \right\} .
\end{eqnarray} We are interested in the $\alpha\to\alpha_c$ limit in
which $\gamma_{\rm w}\to \infty$ and $z=2+\mathcal{O}(\epsilon^2)$.
By first using $x\mapsto x\ell/\sqrt{\gamma_{\rm w}}$ and $t\mapsto
\gamma_{\rm w}\ell^2/x^2$ we transform the two-variable scaling
function into the one-variable white-noise one. Using then
Eq.~(\ref{equA6}) and $A_4 = 2\pi^2$ we obtain the second and third
line below.
\begin{eqnarray} u^2\mathcal{E}^{0,2}(0;\gamma,\gamma_{\rm w}) &=
\frac{2\gamma_{\rm w}g^2A_D(N+2)}{9}\left[\int_{\sqrt{\gamma_{\rm
w}}}^\infty \rmd x\;x^{-3}g_{C_0}^3(0,x^2) \right.  \nonumber\\ &
\left.  \qquad\qquad\qquad\qquad\qquad +\int_0^{\sqrt{\gamma_{\rm w}}}
\rmd x\;x^{-3}g_{C_0}^3(0,x^2)\right] \nonumber\\ &=
\frac{2\gamma_{\rm w}u^2(N+2)}{9}\int_0^\infty\rmd x\left[1 -
\rme^{-x^2/4}\right]^3/ x^3 \nonumber\\ &= \frac{\gamma_{\rm
w}u^2(N+2)}{12}\ln \frac{4}{3}
\label{eq:E11}.
\end{eqnarray} Therefore, at the critical point, using the
Wilson-Fisher fixed point value $u^* = 6\epsilon/(N+8) +
\mathcal{O}(\epsilon^2)$~\cite{Zi96}, we find
\begin{equation}\label{equA9} u^2 \mathcal{E}^{0,2}
(0;\gamma,\gamma_{\rm w}) \to {u^*}^2 \gamma_{\rm w} \mathcal{E}_{\rm
w}^{0,2} = \gamma_{\rm w} \ \frac{3(N+2)}{(N+8)^2} \ \ln\frac{4}{3} \
\epsilon^2 + \mathcal{O}(\epsilon^3) .
\end{equation}

We now compute $\mathcal{E}^{1,1}$ in $D=4$. We start from
Eqs.~(\ref{equ121b}) and~(\ref{eq:defE11}).  Using
Eqs.~(\ref{equA15bis}) and (\ref{equA166}) in the limit $\ell\to 0$ we
obtain
\begin{eqnarray}\label{equA10}
u^2\mathcal{E}^{1,1}(0;\gamma,\gamma_{\rm w}) &=
\frac{g^2A_{4}(N+2)\pi}{144}\frac{-\partial}{\partial\ln\ell}\int_{\ell}^\infty\frac{\rmd
x}{x}\frac{1}{(2\pi)^6} \nonumber\\ &= \frac{{u}^2(N+2)}{72}.
\end{eqnarray} Note that the term coming from the differentiation of
$C_0(\vec x,\ell^z)$ in \eref{equ121b} with respect to $\ln\ell$
vanishes in the limit $\ell\to 0$ [use \eref{equA4}]. At the critical
point we obtain
\begin{equation}\label{equA11} {u^*}^2\mathcal{E}^{1,1} =
\frac{{u^*}^2(N+2)}{72} = \frac{N+2}{2(N+8)^2}\ \epsilon^2 +
\mathcal{O}(\epsilon^3).
\end{equation}


\section{Non-equilibrium propagators}
\label{app:neq}

For $\alpha=1$ the Gaussian non-equilibrium propagators read in the
momentum and time domain
\begin{eqnarray} C_0(\vec{p};t,s) =
\frac{1}{p^2+r}\left[\rme^{-(p^2+r)|t-s|/\gamma_{\rm
w}}-\rme^{-(p^2+r)(t+s)/\gamma_{\rm w}}\right], \\ R_0(\vec{p};t,s) =
\Theta(t-s)\rme^{-(p^2+r)(t-s)/\gamma_{\rm w}} .
\end{eqnarray} For generic $\alpha$, however, analogously compact
expressions are not available and our analysis proceeds using Laplace
transforms. In order to determine the response function $R_0$ ---
consistently with the Gaussian approximation --- we start with the
linearized version of the Langevin equation \eref{equ2} in the
presence of an external perturbation $\vec{h}$:
\begin{equation}\label{equA14} \int_{0}^t \rmd t'\;\Gamma(t-t')
\partial_{t'}\vec{\phi}(\vec{x},t') +
(r-\nabla^2)\vec{\phi}(\vec{x},t') = \vec{\zeta}(\vec{x},t)+
\vec{h}(\vec{x},t)
\end{equation}
Calculating the expectation value of both sides with respect to the
distribution of the noise eliminates the vanishing average $\langle
\vec{\zeta}\rangle$. The Laplace transform yields
\begin{equation}
(\lambda\Gamma_\lambda+p^2+r)\langle\vec{\phi}_{\lambda}(\vec{p})\rangle_h
= \vec{h}_\lambda(\vec p)
\end{equation} in momentum space where we used the Dirichlet boundary
condition $\phi(\vec x, t=0) = 0$ [see discussion at the beginning of
Sec.~\ref{sec:neqpr}].  Note that the expectation value of the order
parameter depends on $h$. The response propagator in the Laplace
domain is given by
\begin{eqnarray} R_0(\vec{p};\lambda,\kappa)\delta_{ij} &=
\frac{\delta\langle\phi_{i,\lambda}(\vec{p})\rangle_h}{\delta
h_{j,\kappa}} |_{\vec h=\vec 0}=
\frac{1}{\lambda\Gamma_\lambda+p^2+r}\frac{\delta
h_{i,\lambda}}{\delta h_{j,\kappa}} \nonumber\\ &=
\frac{1}{(\lambda+\kappa)(\lambda\Gamma_\lambda+p^2+r)}\delta_{ij}\label{equA15bis}
.
\end{eqnarray} The last equality follows from the fact that $\delta
h_i(t)/\delta h_j(s) = \delta_{ij}\delta(t-s)$ as a function of time
translates into $\delta_{ij}/(\lambda+\kappa)$ in Laplace space, given
that $\int_0^\infty \rmd t \rmd s\,\rme^{-\lambda t-\kappa
s}\delta(t-s)=1/(\lambda+\kappa)$.

In order to deduce the correlation propagator we start directly from
\eref{equA14} with $\vec{h}=0$ and we consider its Laplace transform:
\begin{equation} \vec{\phi}_\lambda(\vec{p}) =
\frac{\vec{\zeta}_\lambda}{\lambda\Gamma_\lambda+p^2+r}
\end{equation} which yields
\begin{eqnarray} C_0(\vec{p};\lambda,\kappa)\delta_{ij}
&=\langle\phi_{i,\lambda}(\vec{p})\phi_{j,\kappa}(-\vec{p})\rangle =
\frac{\langle\zeta_{i,\lambda}\zeta_{j,\kappa}\rangle}{(\lambda\Gamma_\lambda+p^2+r)(\kappa
\Gamma_\kappa+p^2+r)} \nonumber\\ &=
\frac{\Gamma_\lambda+\Gamma_\kappa}{(\lambda+\kappa)(\lambda\Gamma_\lambda+p^2+r)
(\kappa\Gamma_\kappa+p^2+r)}\delta_{ij}\label{equA16bis} .
\end{eqnarray} In the last line we used the fact that $\int_0^\infty
\rmd t\rmd s\,\rme^{-\lambda t-\kappa s}\Gamma
(t-s)=(\Gamma_\lambda+\Gamma_\kappa)/(\lambda+\kappa)$. The
propagators verify an `initial time FDT'. We see that for $\alpha<1$
$\lim_{\kappa\to\infty}\kappa R_0(\vec{p};\lambda,\kappa) =
R_0(\vec{p},\lambda)$ and $\lim_{\kappa\to\infty}\kappa^2
C_0(\vec{p};\lambda,\kappa)=
\lim_{k\to\infty}\kappa^{1-\alpha}\Gamma_\lambda
R_0(\vec{p},\lambda)$, with $R_0(\vec p,\lambda) =
1/(\lambda\Gamma_\lambda+p^2+r)$.  In the time domain, the second
identity reads
\begin{eqnarray}\label{eq:initialtimeFDT}
  \partial_{t'} C_0(\vec{p};t,t'\rightarrow 0) \sim
{t'}^{\alpha-1}\int_0^t\rmd s \,
\Gamma(t-s)R_0(\vec{p};s,t'\rightarrow0).
\end{eqnarray} To derive this equation we used the convolution theorem
for the Laplace transform $\mathcal L$, that is $\mathcal
L\left[\int_0^t\rmd t'\ f(t-t')g(t')\right](\lambda) = \mathcal
L[f](\lambda)\mathcal L[g](\lambda)$. In order to deduce the scaling
of \eref{eq:initialtimeFDT} with respect to $t'$ one observes that if
$\lambda\mathcal L[f(t)](\lambda) \sim \lambda^a$ for
$\lambda\to\infty$ then $f(t)\sim t^{-a}$ for $t\to 0$.


\end{appendices}


\section*{References} 
\bibliographystyle{iopart-num}
\bibliography{artbib1}

\providecommand{\newblock}{}
\begin{thebibliography}{10}
\expandafter\ifx\csname url\endcsname\relax
  \def\url#1{{\tt #1}}\fi
\expandafter\ifx\csname urlprefix\endcsname\relax\def\urlprefix{URL }\fi
\providecommand{\eprint}[2][]{\url{#2}}

\bibitem{Masiro73}
Martin P~C, Siggia E~D and Rose H~H 1973 {\em Rev. Rev. A\/} {\bf 8} 423

\bibitem{deDo76}
{De Domninicis} C 1976 {\em J. Phys. Colloques\/} {\bf 37} C1--247

\bibitem{Bajawa76}
Bausch R, Janssen H~K and Wagner H 1976 {\em Z. Phys. B\/} {\bf 24} 113

\bibitem{Hahoma74}
Halperin B~I, Hohenberg P~C and Ma S 1974 {\em Physical Review B\/} {\bf 10}
  139

\bibitem{Hahoma76}
Halperin B~I, Hohenberg P~C and Ma S 1976 {\em Physical Review B\/} {\bf 13}
  4119

\bibitem{Hamaho72}
Halperin B~I, , Ma S and Hohenberg P~C 1972 {\em Physical Review Letters\/}
  {\bf 29} 1548

\bibitem{Hoha77}
Hohenberg P~C and Halperin B~I 1977 {\em Rev. Mod. Phys\/} {\bf 49} 435

\bibitem{DDPe78}
{De~Domninicis} C and Peliti L 1978 {\em Phys. Rev. B\/} {\bf 18} 353

\bibitem{On02}
Onuki A 2002 {\em Phase transition dynamics\/} (Cambridge University Press,
  Cambridge)

\bibitem{Jascsc88}
Janssen H~K, Schaub B and Schmittmann B 1989 {\em Z. Phys. B\/} {\bf 73} 539

\bibitem{cgk-06}
Calabrese P, Gambassi A and Krzakala F 2006 {\em J. Stat. Mech.\/}  P06016

\bibitem{cg-07}
Calabrese P and Gambassi A 2007 {\em J. Stat. Mech.\/}  P01001

\bibitem{Jan92}
Janssen H~K 1992 {\em From Phase Transitions to Chaos---Topics in Modern
  Statistical Physics\/} ed Gy\"orgyi G, Kondor I, Sasv\`ari L and T\'el T
  (World Scientific, Singapore) p~68

\bibitem{cg-05}
Calabrese P and Gambassi A 2005 {\em J. Phys. A\/} {\bf 38} R133

\bibitem{Ozit07}
Ozeki Y and Ito N 2007 {\em J. Phys. A\/} {\bf 40} R149

\bibitem{Zw73}
Zwanzig R 1973 {\em J. Stat. Phys.\/} {\bf 9} 215

\bibitem{Ka73}
Kawasaki K 1973 {\em J. Phys. A\/} {\bf 6} 1289

\bibitem{We08}
Weiss U 2008 {\em Quantum dissipative systems\/} (Singapore, New Jersey,
  London, Hong Kong: World Scientific Publishing Co.)

\bibitem{Hantabo90}
H{\"a}nggi P, Talkner P and Borkovec M 1990 {\em Rev. Mod. Phys.\/} {\bf 62}
  251

\bibitem{Han94}
H{\"a}nggi P 1994 {\em Chem. Phys.\/} {\bf 180} 157

\bibitem{Maweli87}
Masoliver J, West B~J and Lindenberg K 1986 {\em Phys. Rev. A\/} {\bf 34} 1481

\bibitem{SeDi-87}
Seifert U and Dietrich S 1987 {\em Europhys. Lett.\/} {\bf 3} 593--600

\bibitem{Mehwmazh89}
Medina E, Hwa T, Kardar M and Zhang Y~C 1989 {\em Phys. Rev. A\/} {\bf 39} 6

\bibitem{Jafrta99}
Janssen H~K, Frey E and T\"auber U~C 1999 {\em Eur. Phys. J. B\/} {\bf 9} 491

\bibitem{Ka03}
Katzav E 2003 {\em Phys. Rev. E\/} {\bf 68} 046113

\bibitem{Mandel68}
Mandelbrot B~B and van Ness J~W 1968 {\em SIAM Review\/} {\bf 10} 4

\bibitem{Zo09}
Zoia A, Rosso A and Majumdar S~N 2009 {\em Phys. Rev. Lett.\/} {\bf 102} 120602

\bibitem{Po03}
Pottier N 2003 {\em Physica A\/} {\bf 317} 371

\bibitem{Hanju95}
H{\"a}nggi P and Jung P 1995 {\em Advances in Chemical Physics\/} vol~89 ed
  Prigogine I and Rice S~A (John Wiley and Sons) p 239

\bibitem{Cukupe97}
Cugliandolo L~F, Kurchan J and Peliti L 1997 {\em Phys. Rev. E\/} {\bf 55} 3898

\bibitem{Cuku00}
Cugliandolo L~F and Kurchan J 2000 {\em J. Phys. Soc. Japan (Supplement A)\/}
  {\bf 69} 247

\bibitem{GL00}
Godr\`eche C and Luck J~M 2000 {\em J. Phys. A\/} {\bf 33} 9141

\bibitem{Coliza}
Corberi F, Lippiello E and Zannetti M 2007 {\em J. Stat. Mech.\/}  P07002

\bibitem{Cu11}
Cugliandolo L~F 2011 {\em J. Phys. A: Math. Theor.\/} {\bf 44} 483001

\bibitem{Arbicu10}
Aron C, Biroli G and Cugliandolo L~F 2010 {\em J. Stat. Mech.\/}  P11018

\bibitem{Caga02}
Calabrese P and Gambassi A 2002 {\em Phys. Rev. E\/} {\bf 66} 066101

\bibitem{Zi96}
Zinn-Justin J 1996 {\em Quantum Field Theory and Critical Phenomena\/} (Oxford:
  Clarendon Press)

\bibitem{Pa}
Parisi G 1988 {\em Statistical Field Theory\/} (New York: Addison Wesley)

\bibitem{Cu-02}
Cugliandolo L~F 2003 Course 7: Dynamics of glassy systems {\em Slow Relaxations
  and nonequilibrium dynamics in condensed matter\/} ({\em Les Houches\/}
  vol~77) ed Barrat J~L, Feigelman M, Kurchan J and Dalibard J (Springer
  Berlin/Heidelberg) pp 161--171

\bibitem{Ma76}
Ma S~K 1976 {\em Modern Theory of critical phenomena\/} (Benjamin Reading)

\bibitem{LeBe91}
{Le Bellac} M 1991 {\em Quantum and Statistical Field Theory\/} (Oxford: Oxford
  University Press)

\bibitem{Didiei83}
Diehl H~W, Dietrich S and Eisenriegler E 1983 {\em Phys. Rev. B\/} {\bf 27}
  2937

\bibitem{Hamasa09}
Haubold H, Mathai A and Saxena R 2011 {\em J. App. Math.\/} {\bf 2011} 298628

\bibitem{CaGa04}
Calabrese P and Gambassi A 2004 {\em J. Stat. Mech.\/}  P07013

\bibitem{Crri03}
Crisanti A and Ritort F 2003 {\em J. Phys. A\/} {\bf 36} R181

\bibitem{CKP94}
Cugliandolo L~F, Kurchan J and Parisi G 1994 {\em J. Phys. I\/} {\bf 4} 1641

\bibitem{CaGa02a}
Calabrese P and Gambassi A 2002 {\em Acta Phys. Slov.\/} {\bf 52} 335

\bibitem{CaGa02b}
Calabrese P and Gambassi A 2002 {\em Phys. Rev. E\/} {\bf 65} 066120

\bibitem{Sagagu98}
Sancho J~M, Garc{\'\i}a-Ojalvo J and Guo H 1998 {\em Physica D\/} {\bf 113} 331

\bibitem{Ba11}
Barrat J~L and Rodney D 2011 {\em J. Stat. Phys.\/} {\bf 144} 679

\bibitem{Bocuga11b}
Bonart J, Cugliandolo L~F and Gambassi A 2011 {\em in preparation\/}

\bibitem{Ka05}
Kamenev A 2005 {\em arXiv\/}  cond--mat/0412296

\bibitem{Grscin88}
Grabert H, Schramm P and Ingold G~L 1988 {\em Phys. Rep.\/} {\bf 168} 115

\bibitem{Br02}
Bray A~J 1994 {\em Adv. in Phys.\/} {\bf 43} 357

\bibitem{Siarbrcu07}
Sicilia A, Arenzon J, Bray A~J and Cugliandolo L~F 2007 {\em Phys. Rev. E\/}
  {\bf 76} 061116

\end{thebibliography}

\end{document}